\pdfoutput=1

\documentclass[fleqn,usenatbib]{mnras}

\usepackage{newtxtext,newtxmath}

\usepackage[T1]{fontenc}

\DeclareRobustCommand{\VAN}[3]{#2}
\let\VANthebibliography\thebibliography
\def\thebibliography{\DeclareRobustCommand{\VAN}[3]{##3}\VANthebibliography}

\usepackage{graphicx}	
\usepackage{amsmath}	

\usepackage{color} 
\usepackage{soul} 
\usepackage{acronym}
\usepackage{orcidlink}
\usepackage{xurl}

\title[Galaxy Zoo: Cosmic Dawn]{Galaxy Zoo: Cosmic Dawn -- morphological classifications for over 41\,000 galaxies in the Euclid Deep Field North from the Hawaii Two-0 Cosmic Dawn survey}

\author[J. Pearson et al.]{
James Pearson \orcidlink{0000-0001-8555-8561}$^{1}$\thanks{E-mail: james.pearson@open.ac.uk},
Hugh Dickinson \orcidlink{0000-0003-0475-008X}$^{1}$,
Stephen Serjeant \orcidlink{0000-0002-0517-7943}$^{1}$,
Mike Walmsley \orcidlink{0000-0002-6408-4181}$^{2}$,
Lucy Fortson \orcidlink{0000-0002-1067-8558}$^{3}$,
\newauthor
Sandor Kruk \orcidlink{0000-0001-8010-8879}$^{4}$,
Karen L. Masters \orcidlink{0000-0003-0846-9578}$^{5}$,
Brooke D. Simmons \orcidlink{0000-0001-5882-3323}$^{6}$,
R. J. Smethurst \orcidlink{0000-0001-6417-7196}$^{7}$,
Chris Lintott \orcidlink{0000-0001-5578-359X}$^{7}$,
\newauthor
Lukas Zalesky \orcidlink{0000-0001-5680-2326}$^{8}$,
Conor McPartland \orcidlink{0000-0003-0639-025X}$^{9}$,
John R. Weaver \orcidlink{0000-0003-1614-196X}$^{10}$,
Sune Toft \orcidlink{0000-0003-3631-7176}$^{9}$,
Dave Sanders \orcidlink{0000-0002-1233-9998}$^{8}$,
\newauthor
Nima Chartab \orcidlink{0000-0003-3691-937X}$^{11}$,
Henry Joy McCracken \orcidlink{0000-0002-9489-7765}$^{12}$,
Bahram Mobasher \orcidlink{0000-0001-5846-4404}$^{13}$,
Istvan Szapudi \orcidlink{0000-0003-2274-0301}$^{8}$,
\newauthor
Noah East \orcidlink{0009-0005-6387-4877}$^{14}$,
Wynne Turner \orcidlink{0009-0008-3418-5599}$^{15}$,
Matthew Malkan \orcidlink{0000-0001-6919-1237}$^{14}$,
William J. Pearson \orcidlink{0000-0002-7300-2213}$^{16}$,
\newauthor
Tomotsugu Goto \orcidlink{0000-0002-6821-8669}$^{17}$,
Nagisa Oi \orcidlink{0000-0002-4686-4985}$^{18}$
\\
$^{1}$School of Physical Sciences, The Open University, Milton Keynes, MK7 6AA, UK\\
$^{2}$Dunlap Institute for Astronomy \& Astrophysics, University of Toronto, Toronto, ON M5S 3H4, Canada\\
$^{3}$School of Physics and Astronomy, University of Minnesota, Minneapolis 55455, USA\\
$^{4}$European Space Agency (ESA), European Space Astronomy Centre (ESAC), Camino Bajo del Castillo s/n, 28692, Villaneuva de la Cañada, Madrid, Spain\\
$^{5}$Departments of Physics and Astronomy, Haverford College, 370 Lancaster Avenue, Haverford, Pennsylvania 19041, USA\\
$^{6}$Department of Physics, Lancaster University, Lancaster LA1 4YB, UK\\
$^{7}$Oxford Astrophysics, Department of Physics, University of Oxford, Denys Wilkinson Building, Keble Road, Oxford, OX1 3RH, UK\\
$^{8}$Institute for Astronomy, University of Hawaii, 2680 Woodlawn
Drive, Honolulu, HI 96822, USA\\
$^{9}$Cosmic Dawn Center (DAWN); Niels Bohr Institute, University of Copenhagen, Jagtvej 128, 2200 Copenhagen, Denmark\\
$^{10}$MIT Kavli Institute for Astrophysics and Space Research, Massachusetts Institute of Technology, 70 Vassar St, Cambridge, MA 02139, USA\\
$^{11}$Caltech/IPAC, 1200 E. California Blvd., Pasadena, CA 91125, USA\\
$^{12}$CNRS and Sorbonne Université, UMR 7095, Institut d’Astrophysique de Paris, 98 bis, Boulevard Arago, F-75014 Paris, France\\
$^{13}$Department of Physics and Astronomy, University of California, 900 University Ave, Riverside, CA 92521, USA\\
$^{14}$Department of Physics and Astronomy, UCLA, Los Angeles, CA 90095, USA\\
$^{15}$Department of Astronomy, Ohio State University, Columbus, OH 43210, USA\\
$^{16}$National Centre for Nuclear Research, Pasteura 7, 02-093 Warszawa, Poland\\
$^{17}$Institute of Astronomy, National Tsing Hua University, No. 101, Section 2, Kuang-Fu Road, Hsinchu 30013, Taiwan\\
$^{18}$Space Information Center, Hokkaido Information University, Nishi-Nopporo 59-2, Ebetsu, Hokkaido 069-8585, Japan\\
}

\date{Accepted XXX. Received YYY; in original form ZZZ}

\pubyear{\the\year{}}

\begin{document}
\label{firstpage}
\pagerange{\pageref{firstpage}--\pageref{lastpage}}
\maketitle

\newacro{GZ}{Galaxy Zoo}
\newacro{GZCD}{Galaxy Zoo: Cosmic Dawn}
\newacro{HSC}{Hyper Suprime-Cam}
\newacro{H20}{Hawaii Twenty Square Degree}
\newacro{CFHT}{Canada-France-Hawaii Telescope}
\newacro{EDFN}{Euclid Deep Field North}
\newacro{NEP}{North Ecliptic Pole}
\newacro{LSST}{Legacy Survey of Space and Time}
\newacro{LSB}{low surface brightness}
\newacro{AGN}{active galactic nuclei}
\newacro{EROs}{extremely red objects}
\newacro{NCBJ}{National Centre for Nuclear Research}

\begin{abstract}

We present morphological classifications of over 41\,000 galaxies out to $z_{\rm phot}\sim2.5$ across six square degrees of the Euclid Deep Field North (EDFN) from the Hawaii Twenty Square Degree (H20) survey, a part of the wider Cosmic Dawn survey. Galaxy Zoo citizen scientists play a crucial role in the examination of large astronomical data sets through crowdsourced data mining of extragalactic imaging. This iteration, Galaxy Zoo: Cosmic Dawn (GZCD), saw tens of thousands of volunteers and the deep learning foundation model Zoobot collectively classify objects in ultra-deep multiband Hyper Suprime-Cam (HSC) imaging down to a depth of $m_{HSC-i} = 21.5$. Here, we present the details and general analysis of this iteration, including the use of Zoobot in an active learning cycle to improve both model performance and volunteer experience, as well as the discovery of 51 new gravitational lenses in the EDFN. We also announce the public data release of the classifications for over 45\,000 subjects, including more than 41\,000 galaxies (median $z_{\rm phot}$ of $0.42\pm0.23$), along with their associated image cutouts. This data set provides a valuable opportunity for follow-up imaging of objects in the EDFN as well as acting as a truth set for training deep learning models for application to ground-based surveys like that of the Ultraviolet Near-Infrared Optical Northern Survey (UNIONS) collaboration and the newly operational Vera C. Rubin Observatory.

\end{abstract}

\begin{keywords}
methods: data analysis -- catalogues -- galaxies: evolution -- galaxies: statistics -- galaxies: structure -- galaxies: general
\end{keywords}



\section{Introduction}
\label{sec:introduction}

The visual morphology of galaxies provides a valuable means of studying their structure and evolution over cosmic time. Properties derived from morphology pertain to a galaxy’s history, such as the development of discs and spiral structure \citep{2019MNRAS.487.1808M,2021MNRAS.504.3364L,2023RNAAS...7...35M}, the assembly of bulges and bars and the role they play in quenching \citep{2019ApJ...872...50L,2021MNRAS.507.4389G,2024ApJ...973..129G}, the activity of its central black hole \citep{2024MNRAS.532.2320G}, and any previous or current merging events \citep{2018MNRAS.480.2266M}. Such morphological properties can be cross-correlated with one another and compared to, for example, the rate of star formation within galaxies \citep{2015MNRAS.449..820W,2017MNRAS.468.1850H,2022MNRAS.515.3875P}.
Examining morphology can also reveal rare objects, such as \ac{LSB} galaxies which are inherently difficult to detect \citep{1997PASP..109..745B,1997ARA&A..35..267I}, or galaxies acting as gravitational lenses due to their dark matter haloes distorting the light of higher redshift background galaxies situated almost perfectly along the line of sight \citep[e.g.][]{2010ARA&A..48...87T,2024SSRv..220...23L,2025arXiv250315324E}. 

With the combination of many properties contributing to a galaxy's morphology, a large number of classified galaxies are required to study the effect of each of these in detail. The collection of such large samples of galaxies is now underway with the latest and upcoming large-scale surveys like the \textit{Euclid} survey \citep{2011arXiv1110.3193L,2024arXiv240513491E} and the Vera C. Rubin Observatory's Legacy Survey of Space and Time \citep[LSST;][]{2019ApJ...873..111I}, which are pushing toward deeper imaging at higher redshifts while maintaining wide survey areas, in turn allowing the study of galaxy evolution at earlier epochs.
While the \textit{Euclid} Wide Survey will cover 14\,000 square degrees (more than a third of the sky) 
in both the northern and southern hemispheres, the Rubin LSST Wide Fast Deep (WFD) Survey will cover 18\,000 square degrees of the southern sky repeatedly over 10 years.
Additionally, the Ultraviolet Near-Infrared Optical Northern Survey \citep[UNIONS;][]{2025AJ....170..324G} has been obtaining 
imaging from the combination of the \ac{CFHT}, Pan-STARRS, and Subaru \ac{HSC} across 6250 square degrees of the northern sky, acting as a northern complement to Rubin LSST while overlapping with the sky coverage of \textit{Euclid}.


With these surveys observing anywhere from tens of thousands to many millions of objects, image inspection methods are needed in order to study them in detail. Citizen science, specifically crowdsourced data mining, plays a crucial role in this task, utilising the collective effort of thousands of members of the public (`volunteers') to classify large quantities of data much faster than could be achieved by professional researchers alone. Galaxy Zoo \citep{2008MNRAS.389.1179L} is the longest running project on the Zooniverse citizen science online platform, regularly inviting tens of thousands of volunteers to take part in classifying images of galaxies through a web interface. It has seen many iterations over its 17-year lifespan involving collaborations with a number of surveys \citep[e.g.][]{2013MNRAS.435.2835W,2017MNRAS.464.4420S,2017MNRAS.464.4176W,2022MNRAS.509.3966W}, and their results have resulted in numerous further studies of galaxy morphology \citep[a review is presented in][]{2020IAUS..353..205M}.


An obvious complement to citizen science in image classification is the use of machine learning, especially with the growing prevalence of deep learning algorithms in astronomy, including galaxy morphology \citep[e.g.][]{2023PASA...40....1H,2024A&A...683A..42C,2024MNRAS.529..732E,2024MNRAS.530.1274M,2024ApJ...962..164T,2025AJ....169..121L}. Machine learning can be used to automatically classify images almost instantaneously, but for the desired discrete classifications this first requires training the machine learning model on tens to hundreds of thousands of pre-classified images. Additionally, while such models can be used to classify common features like spiral arms and bulge shapes, they often struggle with features that are complex or rare in their training set such as gravitational lenses, irregular galaxies, and mergers. Meanwhile, volunteers have the additional benefit of being able to identify particularly unusual and interesting objects perhaps never seen before through serendipitous discovery. Nevertheless, the far higher classification rate of machine learning makes it desirable as the data sets of images become larger and larger, and so in recent years Galaxy Zoo has employed its own machine learning model, Zoobot \citep{2023JOSS....8.5312W}, alongside its crowdsourcing effort \citep[e.g.][]{2020MNRAS.491.1554W,2022MNRAS.509.3966W,2022MNRAS.513.1581W,2023MNRAS.526.4768W,2024A&A...689A.274E,2025arXiv250315310E}. Having trained on classifications from previous Galaxy Zoo iterations, this model has been used for pre-classification of the simplest objects as well as part of an active learning cycle to continuously retrain as new volunteer classifications arrive in order to provide volunteers with a curated sample of the more difficult and often more interesting objects. This combination of citizen science and machine learning thus provides deeper engagement for volunteers while also speeding up the rate of classifications to tackle the growing size of astronomical data sets. Zoobot is now considered to be a foundation model for extragalactic research \citep{2024eas..conf.2637W,2025arXiv250315310E} and is being employed beyond Galaxy Zoo to other astronomical tasks \citep[e.g.][]{2024RASTI...3..174P,2025arXiv250315326E,2025A&A...696A.214P}.


In this data release, we present results and data sets from one of the latest Galaxy Zoo iterations, \ac{GZCD}, in which Zoobot-assisted volunteers collectively classify tens of thousands of colour postage stamp images from the multiwavelength Cosmic Dawn survey \citep{2024arXiv240805275E} of the \textit{Euclid} deep fields. In particular, these ultra-deep multiband \ac{HSC} images come from Cosmic Dawn's ongoing \ac{H20} survey \citep{2024arXiv240805296E}, covering the first six square degrees of the \ac{EDFN} in the \ac{NEP}.
These deep fields are some of the darkest areas of the sky, selected by Cosmic Dawn for multiple multiwavelength studies in part as preparation for the ultra-deep photometry and spectroscopy of the \textit{Euclid} mission. As such, classifications made by volunteers can provide a means of selecting the most interesting objects for further examination with \textit{Euclid} imaging and follow-up observations. 
For example, the classifications may help improve our understanding of galaxy evolution over a significant fraction of the age of the Universe through identifying rarer objects like galaxy mergers to investigate the role they play in influencing the star-formation rates of galaxies \citep[the extent of how and when this enhancement occurs is still uncertain;][]{2019A&A...631A..51P,2024ApJ...965...60F,2025MNRAS.538L..31F}.

Compared to most previous Galaxy Zoo iterations, this imaging is significantly deeper, which hence allows for the identification of more \ac{LSB} features, and this iteration also places greater emphasis on searching for (higher-redshift) clumpy galaxies and gravitational lenses (see Section~\ref{subsec:decision-tree}), as well as extremely red objects (see Section~\ref{subsec:red-objects}). 
\ac{HSC} is also a precursor for Rubin LSST, and so the morphological classifications in this data release can be utilised as multiband ground truth sets in training deep learning models for application to both Rubin imaging and the HSC imaging of UNIONS.

\ac{GZCD} ran on the Zooniverse platform for over six months, with tens of thousands of volunteers providing almost four million classifications. After the first third of the images were fully classified, a pre-trained Zoobot was further trained on these and deployed in an active learning cycle with volunteers for the remaining two thirds of the images. A final run of Zoobot was also applied to the entire image data set at the end of the project.


This paper is organised as follows. In Section~\ref{sec:imaging-data} we discuss the imaging data used for this work, detailing the surveys, subject selection \& cutout creation methods, and the metadata associated with each image. In Section~\ref{sec:methods}, further details regarding the project implementation are provided, including the decision tree changes, extra tagging task, the project launch, the use of Zoobot machine learning, and data aggregation at the conclusion of the project.
Results are presented in Section~\ref{sec:results}, including a statistical overview of the various morphological classifications and a comparison between volunteer and Zoobot responses. This is followed by a discussion of the data release catalogues in Section~\ref{sec:discussion} and final conclusions in Section~\ref{sec:conclusion}.

The data release catalogues of morphology classifications, metadata, and data set of PNG/JPG cutouts used in \ac{GZCD} are now publicly available for download from \href{https://doi.org/10.5281/zenodo.17200992}{Zenodo} \citep{pearson_2025_17200992} along with additional documentation. See Section~\ref{sec:discussion} for further details.

\section{Imaging Data}
\label{sec:imaging-data}

\subsection{Cosmic Dawn and Hawaii Twenty Square Degree surveys}
\label{subsec:h20-cosmic-dawn}

The Cosmic Dawn survey covers 59 square degrees of the Euclid Deep and Auxiliary Fields from UV/optical to mid-infrared with the aim to understand how galaxies, black holes and dark matter haloes co-evolve from the epoch of reionization to the present -- more details of this survey can be found in \cite{2024arXiv240805275E}. One facet of this is the ongoing \ac{H20} survey, which is covering the two primary \textit{Euclid} deep calibration fields with $u$-band imaging from the \ac{CFHT} MegaCam instrument and ultra-deep $griz$ imaging from \ac{HSC}, the latter of which is on board the 8.2-metre Subaru telescope on the summit of \mbox{Maunakea} in Hawaii. \ac{H20} aims to push the boundaries of extragalactic astronomy by studying galaxy evolution and co-evolution with dark matter haloes out to a high redshift of $z=7$, through combination with Spitzer Space Telescope infrared imaging and Keck DEIMOS (DEep Imaging Multi-Object Spectrograph) spectroscopy. 
Targeting the entire twenty square degrees of the \ac{EDFN}, H20 HSC imaging has intentionally similar depth to \textit{Euclid}'s own near-infrared imaging in this field, with $5\sigma$ depths of $\sim$27 AB mag \citep{2024arXiv240805296E} accompanied by a 0.168 arcsec per pixel angular resolution. HSC imaging is seeing-limited, with a median seeing FWHM of 0.67 arcsec over several years of imaging \citep{2018PASJ...70S...1M}; variations in the PSF between 0.4 and 1.0 arcsec are observed for the H20 survey \citep[Figure 4 of][]{2024arXiv240805296E}.

The data set used in this iteration of Galaxy Zoo consist of ultra-deep $gri$ \ac{HSC} images that cover the first six square degrees of the \ac{EDFN} observed at full depth by the \ac{H20} survey. For reference, the sky coverage of the 9.37 square degrees of the Pre-Launch (PL) \ac{EDFN} Cosmic Dawn catalogue can be seen in Figure 1 of \cite{2024arXiv240805296E}, and more details of how the H20 data was obtained and processed can be found in that work.

\subsection{Subject selection}
\label{subsec:subject_selection}

For source extraction and flux measurements, the H20 pipeline used photometry routines from \textsc{The Farmer} \citep{2023ascl.soft12016W} open-source package, a modified version of \textsc{The Tractor} \citep{2016ascl.soft04008L} that was used for the COSMOS 2020 catalogue \citep{2022ApJS..258...11W}, which fits simple parametric galaxy surface brightness profiles to derive photometry. It was necessary to use this over other methods like SExtractor due to the crowded field observed everywhere at these depths. As part of this, object detection and segmentation was performed with the Source Extraction and Photometry python library \citep[\textsc{SEP};][]{2016JOSS....1...58B}, while photometric redshifts and physical properties were measured with \textsc{EAZY} \citep{2008ApJ...686.1503B}.

Each subject consists of a deep multi-band cutout image centred on an object identified by the H20 pipeline, limited to those with effective radii $r_{\rm eff} \geq 0.6$ arcsec as the HSC PSF's 0.67 arcsec FWHM  made smaller objects too difficult to classify morphologically. In rare cases (<<1 per cent of all detected sources), the pipeline failed to produce suitable photometric models due to a bright neighbouring object, and so for these sources and point sources the detection-measured major axis was used as a proxy for the effective radius.
By taking into account assumptions regarding the average background noise, estimates of the HSC-$i$ band magnitudes for these sources were obtained that were accurate to within $\sim$1 mag, for use in creating image cutouts (see Section~\ref{subsec:cutout-creation}).

In total, this resulted in $\sim$350\,000 sources down to an apparent magnitude of $m_{{\rm HSC}-i}=25$, however this was then limited to only those with $m_{{\rm HSC}-i}<21.5$ to maximise volunteer engagement, based on past Galaxy Zoo iterations and visual inspection of this data set. This was because dimmer sources were far more numerous, and far noisier at lower resolution, meaning the vast majority of them would simply be classed as smooth and featureless, with the rest of the decision tree unnecessary. Occasionally, the H20 pipeline would let through images that were far too zoomed out, so also excluded were any sources with $r_{\rm eff} \geq 500$ arcsec after confirming that these were indeed all erroneous cutouts. As a result, the final data set for \ac{GZCD} contained 47\,347 subjects.

\begin{figure}
	\includegraphics[width=\columnwidth]{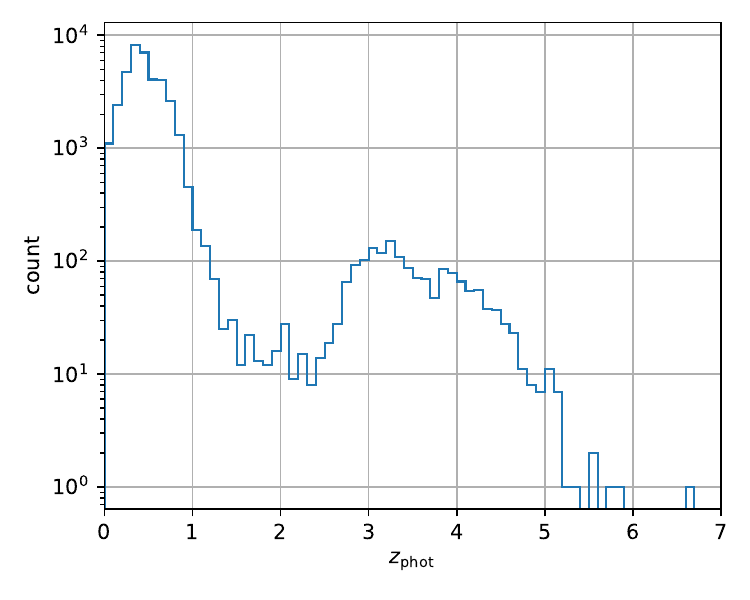}
    \caption{Distribution of photometric redshifts for the objects in \ac{GZCD} as measured by the H20 pipeline. Those for which the pipeline failed to produce suitable photometric redshift estimates are not included. These distributions cover the data set so stars, artifacts and those with `bad image zoom' have not been excluded.}
    \label{fig:z_dist}
\end{figure}

\begin{figure}
	\includegraphics[width=\columnwidth]{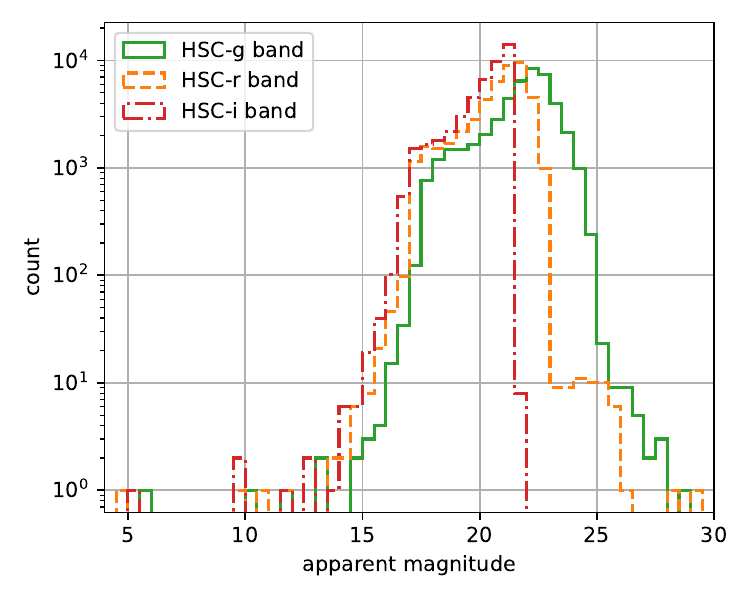}
    \caption{Distribution of HSC-$g$, -$r$, and -$i$ band apparent magnitudes for the objects in \ac{GZCD} as measured by the H20 pipeline. Those for which the pipeline failed to produce suitable photometric models are not included. These distributions cover the data set so stars, artifacts and those with `bad image zoom' have not been excluded.}
    \label{fig:mags_dist}
\end{figure}

The photometric redshift distribution for this subject set can be seen in Figure \ref{fig:z_dist}: while 95.4 per cent have $0 < z_{\rm phot} < 1.0$ (median of $0.42\pm0.23$), the remaining 4.6 per cent extend up to $z_{\rm phot} <= 6.7$. However, it should be noted that for 19.5 per cent of the data set, photometric redshift estimation from the H20 pipeline failed and so these subjects are not included in this range. As discussed in Section~\ref{subsec:aggregated-classifications}, the analysis in later sections is limited to the 45\,742 subjects with $z_{\rm phot} < 2.5$. Figure \ref{fig:mags_dist} shows the distribution of HSC-$g$, -$r$, and -$i$ band apparent magnitudes for the subject set, excluding the small fraction of those for which the modelling failed. Ranges for these magnitudes are \mbox{$5.64<m_{{\rm HSC}-g}<28.81$}, \mbox{$4.95<m_{{\rm HSC}-r}<29.30$}, and \mbox{$5.22<m_{{\rm HSC}-i}<21.5$}, with median averages of 22.03, 21.07, and 20.55, respectively.

\subsection{Cutout creation and upload}
\label{subsec:cutout-creation}

From visual inspection, the cutouts were initially to have widths of approximately 20 times the angular effective radius, however some images still appeared too zoomed in or out. Hence, this was multiplied by a dynamic scale factor\footnote{${\rm max}(a/r_{\rm eff}, b)$, where $a$=(2.0, 1.5, 1.0, 1.0, 0.6) and $b$=(0.6, 0.6, 0.4, 0.3, 0.15) for $m_{{\rm HSC}-i}$<(18, 19, 20, 21, 21.5), respectively. Sources are limited to $r_{\rm eff}\geq0.6$ arcsec so images can only be zoomed out by a factor of 2.0/0.6=3.3 at most. \url{https://github.com/PlanetJames/galaxy-zoo-cosmic-dawn/blob/main/generating_images/GZ_HSC_resizing.py}} that depended on the apparent HSC-$i$ magnitude and effective radius, which for this set of images gave a median average scale factor of $0.8 \pm 0.5$.

\begin{figure}
	\includegraphics[width=\columnwidth]{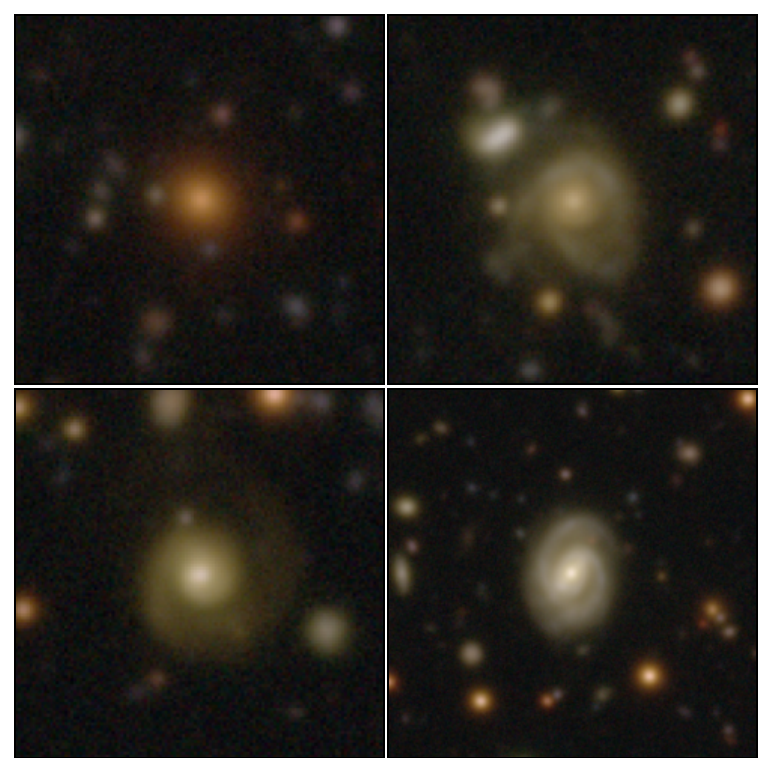}
    \caption{Example cutout images shown to volunteers on Zooniverse as part of \ac{GZCD}. Clockwise from top left, the galaxies have angular effective radii of 1.106, 1.815, 2.483 and 1.441 arcseconds, apparent HSC-$i$ band magnitudes of 19.870, 19.354, 17.356 and 19.150, and photometric redshifts of 0.768, 0.406, 0.301 and 0.398.}
    \label{fig:example-cutouts}
\end{figure}

Colour images were generated using HSC-$g$, -$r$, and -$i$ broad band filters, and colours were balanced used Lupton scaling \citep{2004PASP..116..133L} with parameters based on the HSC Cutout Service\footnote{\url{https://github.com/PlanetJames/galaxy-zoo-cosmic-dawn/blob/main/generating_images/GZHSCScaling.py}}. It was agreed by the Galaxy Zoo team that despite slightly muted colours, this scaling prevented over-saturation and kept a dark background while maintaining low surface brightness features, the latter being a main advantage of this H20 data set. The FITS images were converted to PNG and JPEG images, initially using their native pixel scale of 0.168 arcsec per pixel with 300 DPI before all of these cutouts were resized to 424x424 pixels for upload to Galaxy Zoo. Examples of subjects uploaded are shown in Figure \ref{fig:example-cutouts}.

As the H20 survey observations were ongoing at the time of this Galaxy Zoo iteration, the subject upload process consisted of two main subsets: one batch at project launch, and a second batch later once more observations had been taken and pre-processed by the H20 pipeline.
The first subset contained 16\,671 PNG subjects from the first two square degrees observed by the H20 survey, after limiting the sample according to the selection criteria above: these were to be fully classified by at least 40 volunteers without the aid of machine learning.
The second subset consisted of 30\,676 JPEG subjects from the next four square degrees of the survey: these were to be classified by fewer volunteers with the aid of Galaxy Zoo's deep learning model, Zoobot, in an active learning cycle - see Section~\ref{subsec:zoobot-active-learning}. From initial Zoobot tests, it was decided that the subjects should instead be in JPEG format for faster Zoobot runtimes due to reduced file sizes.

\subsection{Metadata}
\label{subsec:metadata}

Each subject image uploaded to Zooniverse was accompanied by metadata, which can prove useful to researchers and to dedicated volunteers for whom this data increases their engagement with the project. Initially, this primarily consisted of a numerical identifier and name unique to that subject, alongside model parameters from the H20 pipeline for the central object in that image: the angular effective radius (or major axis measured at detection for point-like or failed sources), the apparent HSC-$g$, -$r$, and -$i$ magnitudes, and the median photometric redshift.

Galaxy Zoo has also often provided central coordinates for each subject, in order for more dedicated volunteers to pursue these subjects beyond Galaxy Zoo itself and out to external tools such as virtual observatories of other surveys. However, at the time of launch the H20 data was still private for the benefit of ensuring dedicated H20 publications. After feedback from volunteers, the H20 team wished to help improve the experience and maximise engagement with the project. As such, with permission from the H20 team, the central RA and Dec coordinates were added to the metadata along with an API link to the external H20 Image Viewer\footnote{\url{https://h20.ifa.hawaii.edu/}} (an interactive sky atlas).

\section{Methods}
\label{sec:methods}

\subsection{Decision tree changes}
\label{subsec:decision-tree}

\begin{figure*}
	\includegraphics[width=\textwidth]{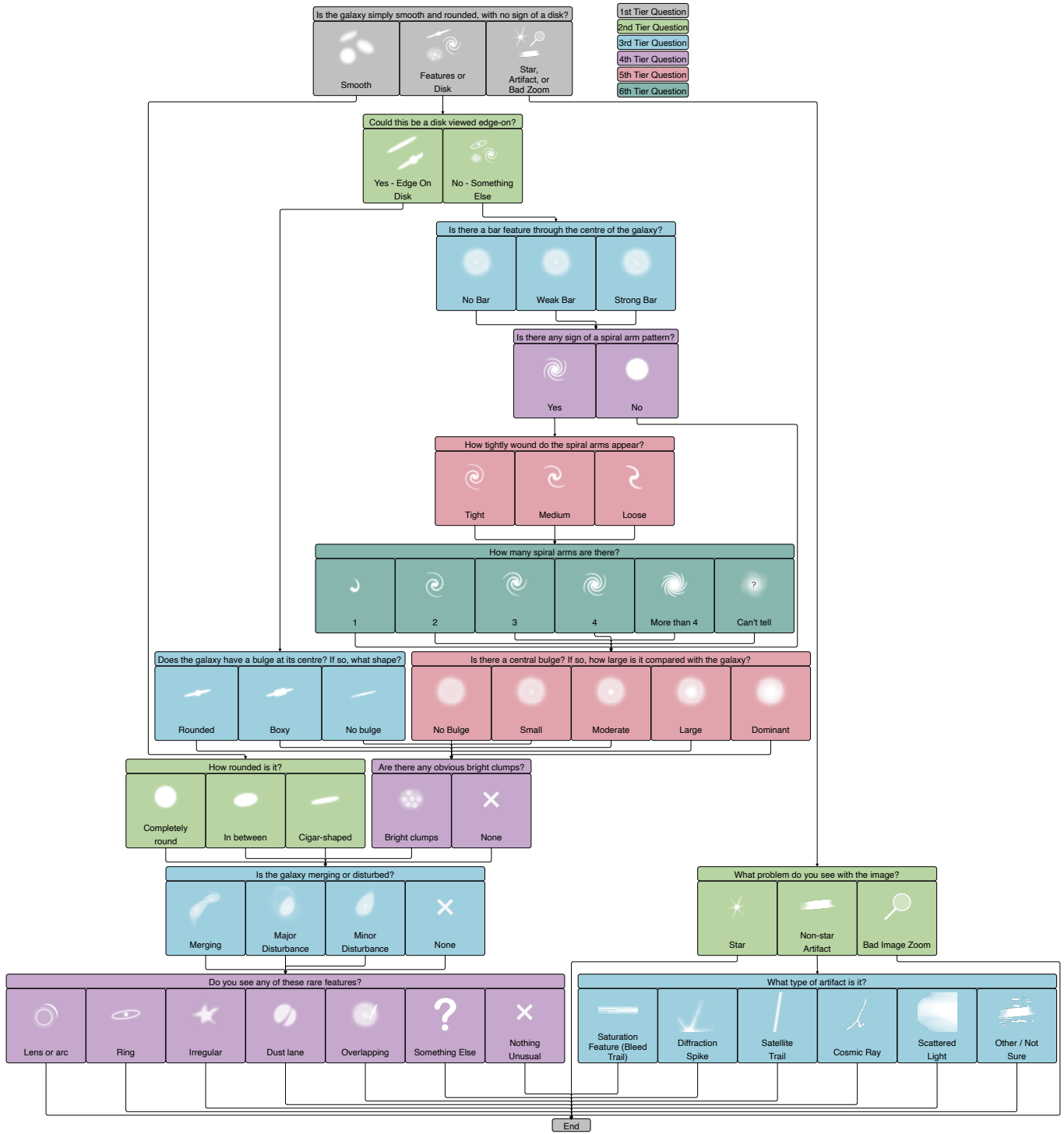}
    \caption{\ac{GZCD} decision tree of questions presented to volunteers for classifying the morphology of the central object in each subject image. Questions sharing the same colour are on the same tier of branching in the decision tree.}
    \label{fig:decision-tree}
\end{figure*}

Galaxy Zoo presents volunteers with one subject at a time, accompanied by a decision tree of questions for them to answer regarding the galaxy at the centre of the subject image. For example, the first question typically asks volunteers to label the galaxy as `Smooth' (featureless), `Features or Disk' (i.e. contains any features or is an edge-on disk), or `Star or Artifact' (i.e. not a galaxy at all). If `Features or Disk' is selected, later questions may ask if there are signs of a spiral arm pattern, and if so, how many spiral arms, and so on. The final question is always the multiple choice question `Do you see any of these rare features?', asking if there are any signs of one or more rare features such as gravitational lensing arcs, ring galaxies, or irregular galaxies. In order for volunteers to fully understand these questions and responses, Galaxy Zoo presents explanations and images within the Zooniverse's built-in Tutorial, Field Guide, and individual `Need some help with this task?' pages.
For this iteration of Galaxy Zoo, a number of changes were made to the decision tree of questions to better suit this data set, which we discuss here. The tutorial, field guide and help pages were also updated to reflect these changes, and the resulting \ac{GZCD} decision tree can be seen in Figure~\ref{fig:decision-tree}.

Firstly, the deep HSC imaging increased the confusion noise of the images, which in rare cases caused the H20 pipeline to treat multiple nearby galaxies as a single source. In these cases, incorrect angular scaling was thus applied to the cutout images, appearing zoomed-out with central galaxies far too small to be classifiable. In order for these subjects to be used for improving the H20 team's reduction pipeline, the `Star or Artifact' response in the Galaxy Zoo decision tree was replaced with `Star, Artifact, or Bad Zoom' (also referred to as `Problem' elsewhere in this work), and selecting this response now presented volunteers with another question rather than moving on to the next subject, asking volunteers which of the three it was: `Star', `Non-star Artifact', or `Bad Image Zoom'.

Secondly, as with all surveys, some artifacts in the H20 images would go undetected in the automated filtering of the reduction pipeline. These could be instrumentation related, such as image ghosts from a nearby bright star, or other sources like cosmic rays, asteroids and satellite trails. To both identify these for possible future study, and to further refine the H20 pipeline, selecting the `Non-star Artifact' responses above then requested volunteers to specify which type of artifact it was, with the following possible responses: `Saturation Feature (Bleed Trail)', `Diffraction Spike', `Satellite Trail', `Cosmic Ray', `Scattered Light', `Ghost', or `Other / Not Sure'. This last option included any other astronomical object that was not a galaxy, such as a nebula or star-forming cloud.


Thirdly, the deep imaging of the H20 survey allowed for resolving galaxies at earlier epochs in the Universe's evolution, where `clumpy' galaxies are thought to be more common \citep{2022ApJ...931...16A}. Unlike the vast majority of galaxies in the nearby Universe, these galaxies have large kpc-scale regions of enhanced star formation that appear as separated bright clumps in images, particularly within spiral arms - around 60 per cent of star-forming galaxies at $z\sim2$ contain these clumps.
These clumps provide a record of a galaxy's past and ongoing star formation, and hence a means of studying how such galaxies form and evolve, to answer questions about what physical mechanisms produce these clumps and why so few clumps exist in the nearby Universe in comparison.
As such, the question `Are there any obvious bright clumps?' (with responses `Yes' and `No') was added to the Galaxy Zoo decision tree to help identify these galaxies.


Finally, the nature of these deep HSC images raised the possibility of detecting higher redshift strong gravitational lenses. Modelling the mass density profile of these rare objects can offer valuable insights into galaxy evolution, constraining cosmological models, and the nature of dark matter and dark energy.
For example, measuring the low-mass substructures within the foreground lens can place limits on the properties of dark matter. Galaxy evolution can be studied through the stellar-to-halo mass relation, comparing the modelled halo masses of lenses with their stellar masses obtained from other methods, and modelling of the lens can also be used to reconstruct the higher redshift lensed source to probe the early Universe. Additionally, time-delay lenses can be modelled to constrain the Hubble constant $H_0$, and double source plane lenses can constrain other cosmological parameters independent of $H_0$.
To best realise these uses for strong lenses requires high-quality lenses of specific types or large statistical samples of lenses, at a range of redshifts, with higher redshift lenses extremely valuable for probing the early Universe.
As such, for this iteration of Galaxy Zoo a greater emphasis was placed on detecting these gravitational lenses, with `Lens or arc' shifted to the top of the list of responses for the `rare features' question, and volunteers also tasked with tagging any lenses they find (see Section~\ref{subsec:red-objects} for details about the Zooniverse tag system).

\subsection{Tagging extremely red objects}
\label{subsec:red-objects}

The Zooniverse provides a built-in method for volunteers to assign keywords (`tags') to subjects they find particularly interesting or possess features not mentioned in the decision tree. After classifying a subject, volunteers can choose to move to the project's Talk boards where they can leave comments about it, including any keywords they deem appropriate - there is no set list of keywords, so as to allow for anything unusual to be labelled. For example, subjects in Galaxy Zoo might be tagged as `\#disturbed', `\#lsb', `\#star-forming', and so on, now also with an emphasis on tagging gravitational lenses with `\#lens' following Section~\ref{subsec:decision-tree}.

One benefit of the tagging system for Galaxy Zoo is for finding objects of a certain colour, as classification by colour is not included in the decision tree. For \ac{GZCD}, the majority of galaxies appeared blue, white, or golden-orange due to the chosen colour scaling, and so the H20 team was particularly interested in identifying any unusually red galaxies appearing in the data set that may be indicative of \ac{EROs}. These objects are at higher redshifts ($z>1$) and typically correspond to either old, evolved stellar populations in elliptical galaxies, or the dust reddening of dusty star-forming galaxies or \ac{AGN}. While in principle the classification of objects based on colour could be automated, the crowded fields seen in deep H20 images can make this challenging, and such objects could be entirely misclassified by automated algorithms if they are overlapping or otherwise too near other sources. The tagging of potential \ac{EROs} can hence also aid in improving the H20 pipeline's automated algorithms.

As such, volunteers were tasked with tagging these rare objects with either `\#bright-red' or `\#dim-red', through instructions provided in the Galaxy Zoo tutorial, help guide, and Talk Board. As with the decision tree, volunteers were asked to focus only on the central object in order to avoid duplication of classifications (as they may have appeared as central galaxies in other cutouts) or the tagging of red stars (which had likely already been filtered out of being central objects by the H20 pipeline). However, to account for any potential errors with the H20 pipeline, volunteers were allowed to tag subjects in this way if they contained \ac{EROs} that were overlapping or merging with the central galaxy, or acting as a lensing arc. Additionally, to ensure completeness but avoid confusion, it was recommended that any \ac{EROs} notably away from the centre could still be tagged using `\#bright-red-companion' or `\#dim-red-companion'.

\subsection{Public launch}
\label{subsec:public-launch}

\ac{GZCD} launched on 21 October 2022, accompanied by a post on the Galaxy Zoo Talk boards, a post on the Galaxy Zoo Blog\footnote{\url{https://blog.galaxyzoo.org/2022/10/21/the-dawn-of-galaxy-zoos-new-incarnation-galaxy-zoo-cosmic-dawn/}}, an advert in the Zooniverse email newsletter to Zooniverse account holders, and additionally shared on social media. 
For further promotion early in the project, there were also multiple press releases from The Cosmic Dawn Centre, The Open University, the \ac{NCBJ}, and the Science in Poland website, with the latter two also translated into Polish.

It was highlighted to volunteers that, compared to most previous iterations of Galaxy Zoo, the Cosmic Dawn data set has significantly deeper imaging, containing more distant galaxies from earlier in the Universe’s history and covering a new area of the sky which had not been targeted by such deep observations before. As a result, the volunteer experience was a little different given the lack of much additional multiwavelength data from other surveys for them to explore with external tools, but it was emphasised to volunteers that the vast majority of subjects are brand new, so they would have real chance of making exciting new discoveries - their eyes would likely be the first to see many of the galaxies they are classifying.

\subsection{Zoobot active learning}
\label{subsec:zoobot-active-learning}

After the full classification of the first subset of 16\,671 subjects by at least 40 volunteers per subject (as mentioned in Section~\ref{subsec:cutout-creation}), these were used to fine-tune Galaxy Zoo's deep learning model, Zoobot. The model had already been trained on previous iterations of Galaxy Zoo, and this fine-tuning would improve its ability to handle HSC images and similar deep imaging in future.

For this, Zoobot used an \textsc{EfficientNetB0} architecture \citep{2019arXiv190511946T} finetuned from the `GZ Evo' data set \citep{2024arXiv240402973W}, which consisted of subjects from the following previous Galaxy Zoo iterations: Galaxy Zoo 2 (GZ2), GZ Hubble, GZ CANDELS, GZ DECaLS, GZ DESI and GZ Rings, where CANDELS, DECaLS and DESI correspond to the Cosmic Assembly Near-infrared Deep Extragalactic Legacy Survey, Dark Energy Camera Legacy Survey and the Dark Energy Spectroscopic Instrument, respectively.
The model outputs predictions for the fractions of volunteers who would select (`vote' for) each response to each decision tree question across decision trees from all Galaxy Zoo iterations, except for the `rare features'
question due to it uniquely allowing multiple responses to be selected. Specifically, the model predicts Dirichlet distributions for the possible fractions, from which we calculate the most likely (mean) fraction as well as the uncertainty (upper and lower 90 per cent confidence intervals) on the possible fractions.

With the start of classifications for the second subset of 30\,676 subjects, Zoobot was deployed alongside volunteers in an active learning cycle. First, the images were passed through Zoobot, and a retirement threshold was set such that a subject would be retired (i.e. prevented from receiving any further volunteer classifications) if Zoobot predicted with 90 per cent confidence that fewer than 20 per cent of volunteers (vote fraction < 0.2) would select the `Features or Disk' response to the first question, regardless of predicted responses for the other decision tree questions. The remaining subjects (i.e. galaxies likely to have features) would then be shown to volunteers for classification. Each week, all volunteer classifications collected to date -- including the new classifications from the previous week -- were used to retrain Zoobot, which was then applied to the images once again, and so on. Zoobot is applied to all unretired subjects, and so can `change its mind' (i.e. move galaxies between the active featured set and the hidden non-featured set) as performance improves.

The use of active learning serves two main benefits. First, it can greatly speed up the classification process, requiring fewer volunteer classifications per subject. Second, the deep learning model typically quickly retires smooth elliptical galaxies that volunteers would find uninteresting, leaving volunteers to focus on more visually complex objects that they would find more interesting in order to maximise engagement.

Initial testing suggested this active learning would retire $\sim$77 per cent of galaxies early and speed up classification by a factor of $\sim$5.
Later evaluation showed that Zoobot allowed the majority of subjects to be confidently retired with less than 40 volunteer classifications: the model did indeed retire 77.8 per cent of galaxies early, and 528 of these subjects did not receive any volunteer classifications as Zoobot was confident enough in its initial classifications to immediately retire them before they could be seen by volunteers. 
Additionally, the majority of subjects retired early were smooth galaxies, as expected: of the subjects that received less than five volunteer votes before retirement, 62 per cent were voted as `Smooth' by the majority of volunteers (95 per cent by Zoobot for the 528 Zoobot-only subjects), compared to only 3 per cent (0 per cent) and 12 per cent (2 per cent) for `Features or Disk' and `Problem', respectively.
Through confident classifications from the model leading to early subject retirement, the average rate of classifications increased by a factor $\sim$2.4 for the second subset, with the 30676 subjects retired over 83 days compared to the 16671 subjects of the first subset that were retired over 109 days. It should be noted that this is a lower estimate for the speed-up that Zoobot active learning can provide, given the higher numbers of volunteers taking part during the first subset. 
To account for this, the speed-up can alternatively be estimated to be a factor of $\sim$2.9 given that subjects in the second subset only required a mean average of 14 volunteer classifications before retirement instead of 40. Additionally, the 77.8 per cent retired early by Zoobot were skewed towards a mean average of only 6 volunteer classifications (median of 4), for which the speed-up was therefore a factor of $\sim$7.

\subsection{Data aggregation}
\label{subsec:data-aggregation}

After the completion of \ac{GZCD}, the classification data was aggregated to provide total and fractional votes (classifications) for each response to each decision tree question for each subject. Final Zoobot classifications were also obtained (see Section~\ref{subsec:zoobot-classifications}) and tags for each subject were collected through the Zooniverse interface. See Section~\ref{sec:discussion} for details on the final catalogues released with this paper.

\subsubsection{Volunteer weightings and redshift debiasing}
\label{subsec:weightings-and-debiasing}

In some previous iterations of Galaxy Zoo, post-processing may be applied to the `raw' vote fractions in order to improve results. One method is through weighting each volunteer's responses according to how well the volunteer generally agrees with the consensus. These weighted vote fractions are calculated by iteratively re-weighting each user based on their general agreement with other users, although the effect on performance is typically minor \citep[e.g.][]{2017MNRAS.464.4176W}. 
Additionally, given the limited resolution of the images shown to volunteers, vote fractions inevitably reflect the appearance of the galaxy, not its true underlying morphology, such that higher redshift objects are more likely to be classified as `Smooth' rather than `Features or Disk'.
Hence in previous works, attempts have been made to `de-bias' the vote fractions, estimating how objects would have been classified at lower redshift and hence their true morphology rather than that presented to volunteers \citep{2016MNRAS.461.3663H}. A more comprehensive explanation and reasoning for Galaxy Zoo weightings and debiasing efforts can be found in Section\,4.3 of \cite{2022MNRAS.509.3966W}, which recommends only using debiased classifications to consider populations of galaxies rather than individuals.
In this work, no volunteer weighting or debiasing was applied to the results, for the reasons we will expand below.

For weighting, as per \cite{2017MNRAS.464.4176W}, only logged-in users were considered, as weighting accounts for users who repeatedly classify incorrectly, while still allowing for disagreements between volunteers that can lead to serendipitous discovery. Following \cite{2013MNRAS.435.2835W}, a volunteer's weighting $w$ is calculated as follows
\begin{equation}
    w = \min(1.0, (\overline{k}/0.6)^{8.5})
	\label{eq:volunteer_weighting}
\end{equation}
where $\overline{k}$ is the mean average consistency of that volunteers' votes compared to those of other volunteers. However, for this work it was found that only 0.8 per cent of logged in volunteers would have a weight less than 1.0 following $\overline{k}$ values below 0.6. Previous works \citep{2017MNRAS.464.4420S,2017MNRAS.464.4176W,2022MNRAS.509.3966W} have highlighted how the main change volunteer weighting brings is to eliminate the small percentage of volunteers who each have hundreds of classifications containing a high proportion of responses labelling objects as `Star or Artifact'. This is thought to be due to previous Galaxy Zoo iterations having such a response to the first question lead directly to the next subject, allowing these volunteers to skip through many subjects quickly by performing mostly inaccurate classifications, whether that be to view more subjects or to artificially increase their classification count. However, this Cosmic Dawn iteration expands the decision tree so that the updated `Star, Artifact or Bad Zoom' response now leads to further questions, a by-product of which was found to be the elimination of this behaviour: no cases of this were found in the aggregated classifications. Based on this, alongside the minor number of volunteers that would be affected by weighting, it was deemed unnecessary for the volunteers' vote fractions to be weighted.

For debiasing vote fractions, it should be noted that the current debiasing method assumes galaxy morphological properties do not evolve significantly for those of similar intrinsic brightness and physical size. While some previous iterations of Galaxy Zoo have involved sets of objects sharing similar, and often low, redshifts, for which this assumption approximately holds \citep{2022MNRAS.509.3966W}, the Cosmic Dawn iteration's subject set covers a much wider redshift range (primarily $0 < z < 1.0$, as well as many objects at higher redshift). Additionally, photometric redshift estimation from the H20 pipeline failed for approximately one fifth of the sources in the subject set. For these reasons, no redshift-based debiasing was applied to the results, with the final vote fractions remaining `raw', i.e. entirely based on the visual morphology present in the images shown to volunteers. As such, classifications of objects at higher redshifts will be more adversely affected due to lower resolution, with them being more likely to be classified as non-featured \citep[for more information on the effect of this biasing, see][]{2023MNRAS.526.4768W}.

\subsubsection{Galaxy Zoo Mobile gravitational lenses}
\label{subsec:lenses}

Throughout the project, in addition to the `rare features' question, volunteers were asked to tag any gravitational lenses they found on the Talk boards: given the rarity of such objects, we wished to explore all possible methods for compiling as complete a sample as possible, especially given that the `rare-features' question is shown last to volunteers. As the project neared completion, another method was explored: in addition to the desktop version of Zooniverse, there is also a Zooniverse app for mobile devices, intended for simple and often binary `Yes/No' workflows that are compatible with the mobile app's `swipe left/right' functionality. Hence, as a follow-up to the project, a workflow for finding gravitational lenses through the mobile app was launched alongside other workflows as part of a reboot of Galaxy Zoo Mobile. It consisted of 22\,587 subjects, which were simply the set of galaxies that had not been fully classified in the desktop application, and removing those predicted by Zoobot to be stars, artifacts or images with `Bad Image Zoom' with a 90 per cent lower bound greater than 0.5.
The workflow asked volunteers if each image contained a gravitational lens or not, and used a retirement threshold of ten volunteer classifications given the far simpler, binary `Yes/No' workflow. To improve speed and volunteer experience, an additional retirement rule was introduced such that a subject would be retired if it had six or more votes and every one of those votes was for the `no' response.

\section{Results}
\label{sec:results}

\subsection{Volunteer classifications}
\label{subsec:volunteer-classifications}

\subsubsection{Volunteer \& classification number counts}
\label{subsec:number-counts}

\begin{figure}
	\includegraphics[width=\columnwidth]{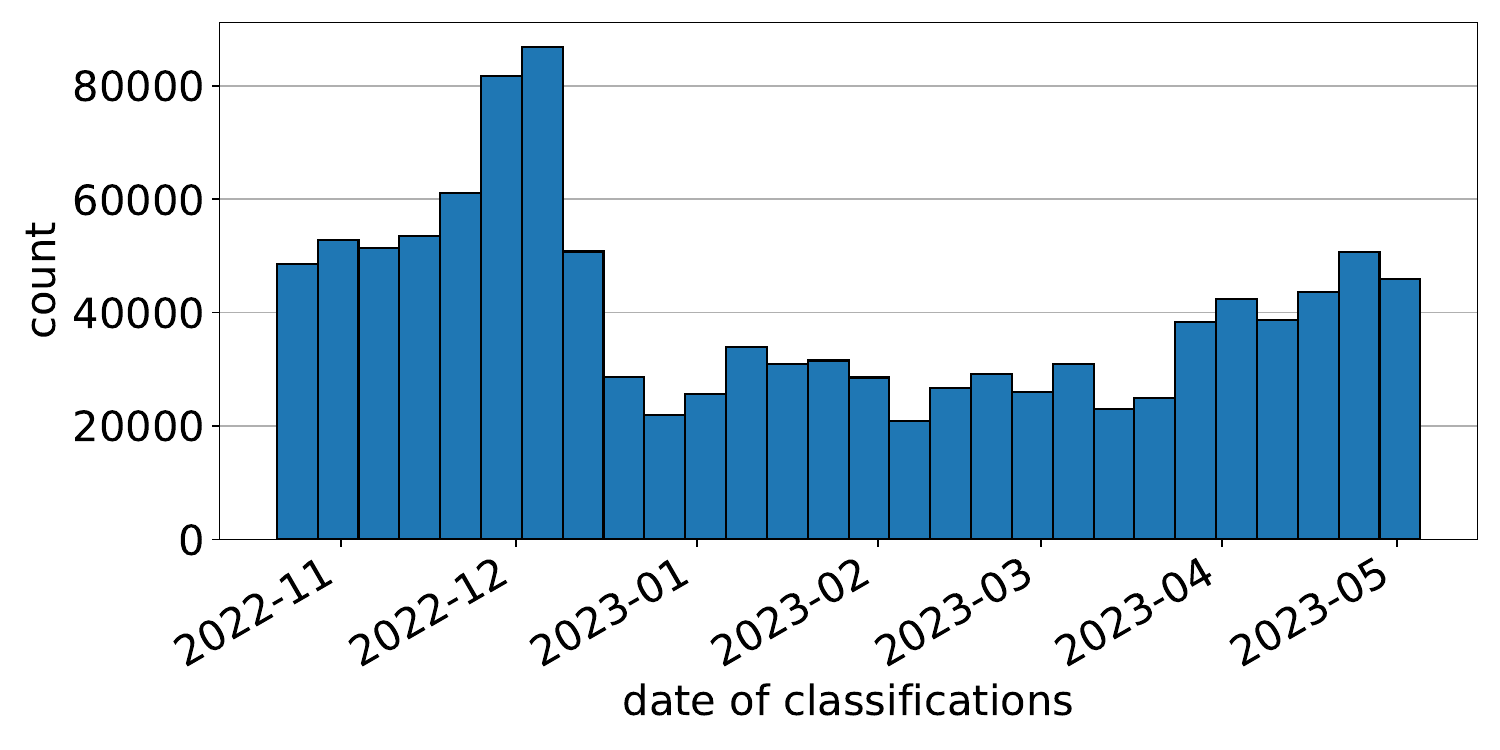}
    \caption{Histogram of the number of volunteer classifications (`votes') made per week across the whole run of \ac{GZCD}.}
    \label{fig:classifications-per-week}
\end{figure}

\ac{GZCD} ran for over six months from 21 October 2022 to 03 May 2023 and consisted of 47\,347 subjects. 3\,990\,911 classifications were made by tens of thousands of volunteers: 74 per cent by the 10\,274 logged-in volunteers along with 26 per cent from an unknown number of volunteers who made classifications without logging in. 
The weekly number of total classifications made by volunteers are provided in Figure \ref{fig:classifications-per-week}. The peak at the end of November corresponds to an update put out on the Galaxy Zoo Talk boards, in particular highlighting the release of the H20 Image Viewer (see Section~\ref{subsec:metadata}) and, correspondingly, the inclusion of image coordinates in the Galaxy Zoo subject metadata to view them beyond the Zooniverse platform. 
Unfortunately, the press releases (see Section~\ref{subsec:public-launch}) at the end of 2022 did little to increase volunteer engagement, perhaps due to a limited ability to reach would-be volunteers compared to, say, the post on the Galaxy Zoo Blog at project launch. The rise in classifications towards the end of the project was likely due to volunteers becoming more engaged as the `percentage complete' bar nears completion in addition to seeing more interesting images as a result of Zoobot active learning.

\begin{figure}
	\includegraphics[width=\columnwidth]{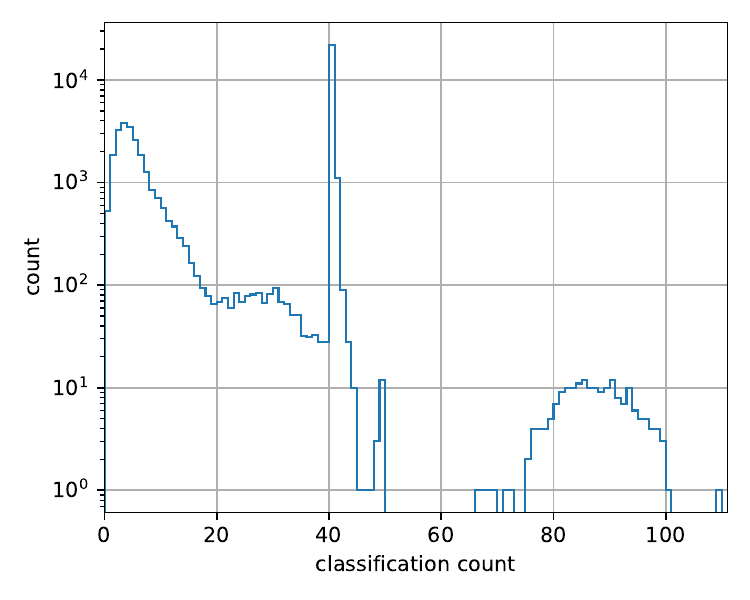}
    \caption{Histogram of the number of volunteer classifications (`votes') made per subject across the whole data set. The peak at 40 classifications corresponds to the threshold for retiring a subject.}
    \label{fig:classification-count}
\end{figure}

The first set of 16\,671 subjects was fully classified by at least 40 volunteers per subject. As discussed in Section~\ref{subsec:zoobot-active-learning}, with the introduction of Zoobot active learning to the second set, 528 of the subjects were immediately and confidently retired by Zoobot without any volunteer classifications; the remaining 30\,148 each received at least one volunteer classification, of which 19\,686 (65 per cent) had fewer than ten volunteer responses before being retired by Zoobot.
Figure \ref{fig:classification-count} shows the number of votes that each subject received from volunteers. For those with less than 40 votes, there is a skew towards low counts, due to Zoobot retiring many subjects quickly and more over time as it trains through active learning. The threshold for retirement is made obvious by the peak at 40 votes, which, in addition to subjects in the first set each receiving at least 40 votes, is amplified by 22.6 per cent of subjects in the second set also receiving at least 40 votes: for these cases, if Zoobot could not confidently classify a subject after being seen by a small number of volunteers, it was unlikely to ever retire it early, leading to those subjects hitting (or exceeding) the retirement threshold. 
There are many subjects that received slightly more than 40 votes, as those with just under 40 votes could receive classifications from multiple volunteers almost simultaneously before the automated retirement could be enacted.

Figure \ref{fig:classification-count} also shows another peak in the distribution at higher classification counts, with 189 subjects (0.4 per cent of data set) receiving 66-109 votes each, with a median of 87 votes.
Of these, 100 have randomly distributed H20 IDs: when a subject set is completed (with all subjects having reached one or more retirement criteria), Zooniverse continues to show a sample of 100 subjects, which are still classifiable but marked by a `Finished!' tag, which aids in maintaining volunteer engagement before a new set becomes available.
The other 89 subjects correspond to the final IDs in the first set, which all received a significant fraction of their votes ($\sim$48-93 per cent) towards the end of that set's classification period. Why these subjects received so many classifications is uncertain, but likely due to an unknown bug with the Zooniverse platform during the course of the project runtime: they do not correlate to any particular object and are not duplicate uploads, nor are they erroneously duplicated entries in the classification results, and the vast majority of the votes come from logged-in volunteers so cannot be due to bot accounts repeatedly classifying the same subjects.

\begin{figure}
	\includegraphics[width=\columnwidth]{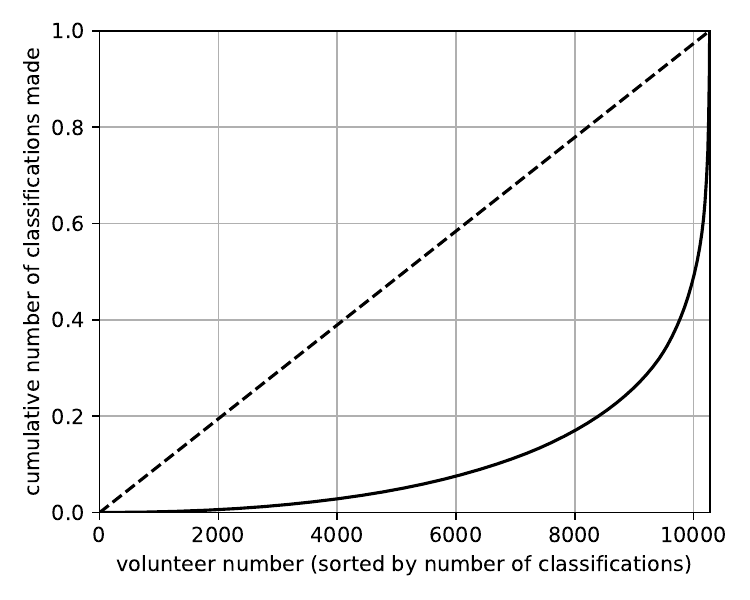}
    \caption{A Lorenz curve depicting the cumulative number of classifications made by each logged-in volunteer, ordered from least to most contributions and normalised by the total number of logged-in classifications. The dashed line indicates where the curve would lie if all volunteers contributed the same number of classifications.}
    \label{fig:classifications-per-volunteer}
\end{figure}

The cumulative number of classifications made by each logged-in volunteer are shown in Figure \ref{fig:classifications-per-volunteer}, as a Lorenz curve in the same style as for Galaxy Zoo CANDELS \citep{2017MNRAS.464.4420S}. Such a curve can be quantified by the Gini coefficient, where values of 0 and 1 correspond to classifications being equally shared amongst volunteers or entirely performed by one volunteer, respectively -- examples of Gini coefficients from various Zooniverse projects can be found in \cite{2029JCOMSpiers}. Compared to a value of 0.86 from Galaxy Zoo CANDELS (which featured a similar number of subjects), the Gini coefficient for \ac{GZCD} is 0.79 for logged-in classifications, and hence such classifications for this iteration are spread among a greater fraction of volunteers, perhaps due to an increase in engagement with subjects from Zoobot active learning. This is reflected in the fact that the top 18 per cent of \ac{GZCD} volunteers contributed 80 per cent of the logged-in classifications, compared to the top 9 per cent in Galaxy Zoo CANDELS, although this does not translate to a greater overall number of volunteers as the latter iteration involved around ten times more volunteers.

\subsubsection{Aggregated classifications}
\label{subsec:aggregated-classifications}

\begin{figure}
	\includegraphics[width=\columnwidth]{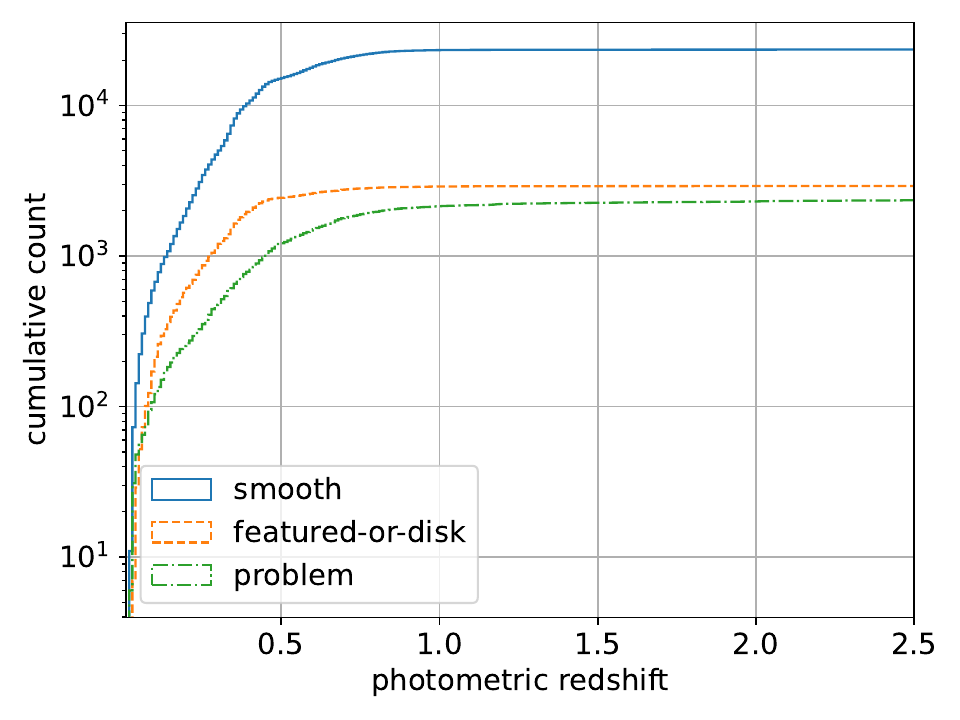}
    \caption{Cumulative number counts for each primary response of the decision tree, as a function of photometric redshift. Those included here were voted as that response by more than 50 per cent of volunteers, however those without photometric redshifts have been excluded.}
    \label{fig:morphology-vs-z}
\end{figure}

In agreement with the H20 Cosmic Dawn team, the results of aggregated votes presented in this work, alongside the released classifications and cutout images, are limited to the 45\,742 subjects with either $z_{\rm phot} < 2.5$ or for which the photometric redshift estimation failed, such that follow-up work on sources with higher estimated redshifts (3.4 per cent of the data set) can be carried out by the H20 Cosmic Dawn team themselves.
For this redshift-limited sample, the cumulative number counts for each of the first responses -- `Smooth', `Features or Disk', and `Star, Artifact, or Bad Zoom' (referred to here as `Problem') -- are given as a function of photometric redshift in Figure \ref{fig:morphology-vs-z}. Limited to those with photometric redshifts as per the figure, the total number of subjects voted by more than 50 per cent of volunteers as being `Smooth', `Features or Disk', or `Problem' subjects were 23\,533, 2920, and 2346, respectively, while the remaining 7282 subjects' votes were less certain. Including the objects without photometric redshifts, these figures become 28\,605 (63 per cent), 3201 (7 per cent), and 4123 (9 per cent), respectively, with 9297 (21 per cent) subjects' votes less certain.
As such, by separating out the `Problem' images, this indicates the data set contains at least 41\,000 cutouts with central galaxies.
For the sample of subjects with $z_{\rm phot}\gtrsim1.0$, the majority of galaxies are seeing-limited so observationally biased towards being classified as `Smooth'. However, there are still extended objects in this sample, with 14 per cent of subjects classified as `Features or Disk'.

\begin{figure}
	\centering
	\includegraphics[width=0.9\columnwidth]{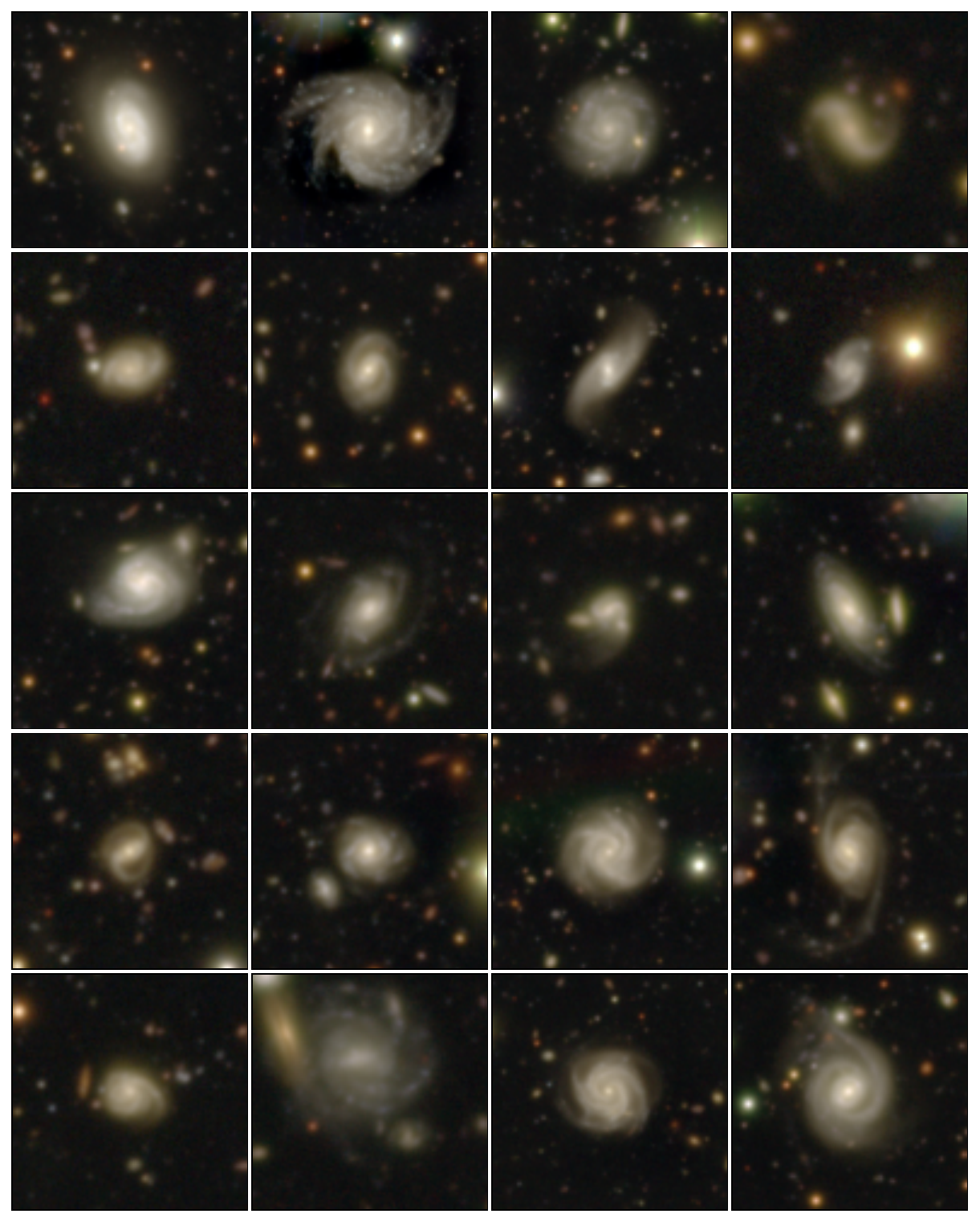}
    \caption{Example \ac{GZCD} subjects voted as being `Features or Disk' in response to the first question, having received the highest volunteer vote fractions for that response. Subjects are limited to those with at least ten volunteer votes.}
    \label{fig:top20-featured}
\end{figure}

\begin{figure}
	\centering
	\includegraphics[width=0.9\columnwidth]{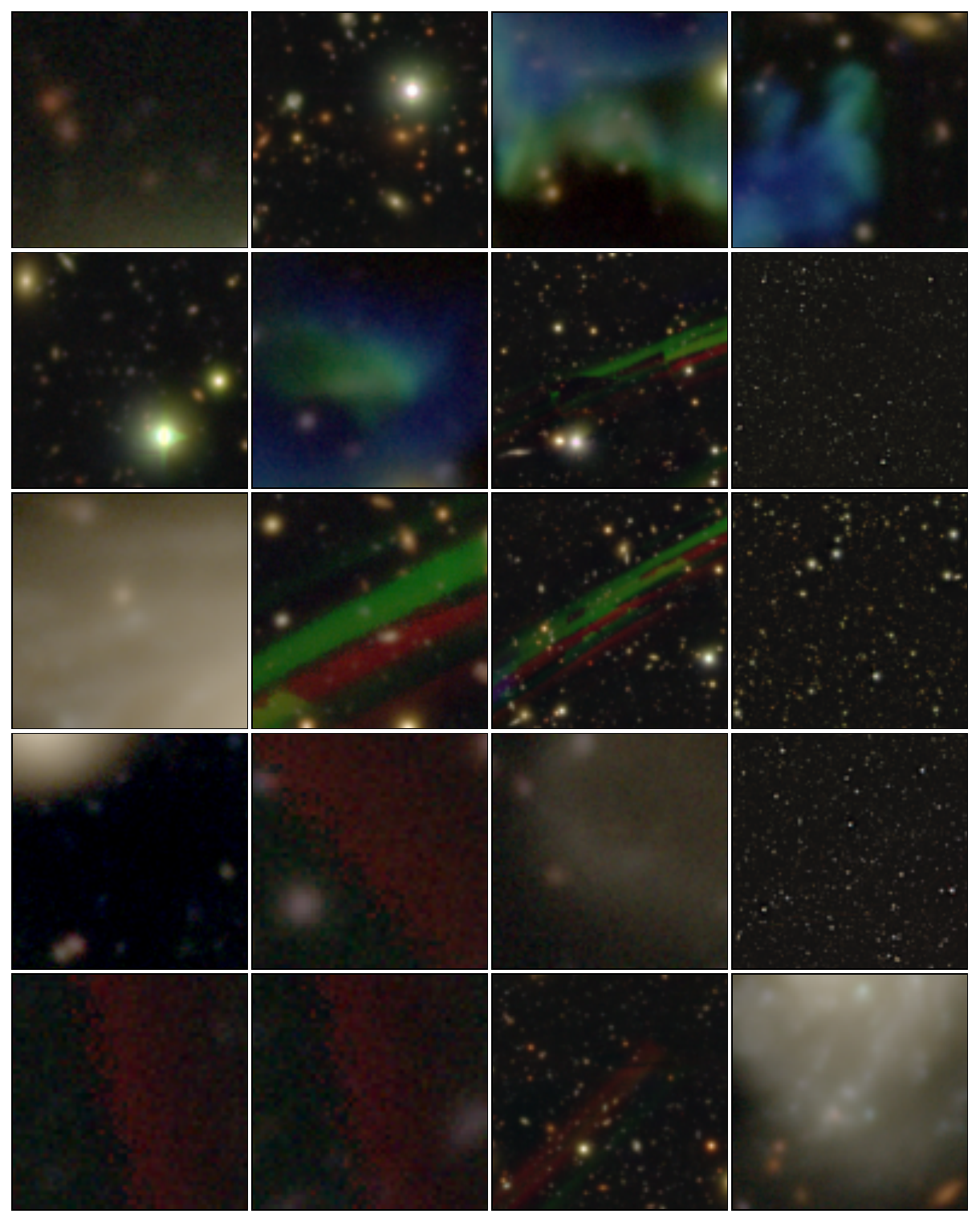}
    \caption{Example \ac{GZCD} subjects voted as being `Star, Artifact, or Bad Zoom' in response to the first question, having received the highest volunteer vote fractions for that response. Subjects are limited to those with at least ten volunteer votes. The hazy green and blue objects in the third, fourth and sixth images correspond to the ejected material surrounding the Cat's Eye Nebula.}
    \label{fig:top20-problem}
\end{figure}

Some examples of classified subjects from the full data set are provided in Figures \ref{fig:top20-featured} and \ref{fig:top20-problem}, which show those with the highest volunteer vote fractions for their response options. 
The first contains subjects with the highest vote fractions for `Features or Disk', one of the responses to the first question presented to volunteers, with the majority of those with high vote fractions for this response containing clear spiral galaxies.
Figure \ref{fig:top20-problem} on the other hand demonstrates an array of `Star, Artifact, or Bad Zoom' subjects in response to the first question, including those featuring the most common image artifacts, which are then further classified into separate types of artifact in a follow-up question. 
From visual inspection of those voted as `Bad Image Zoom', almost all were either zoomed-in portions of a nearby galaxy, or very zoomed out to the point of having no notable central galaxy. Regardless, these cutouts did not undergo reprocessing as any galaxies in them were also already present as correct central galaxies in other cutouts.
While over a thousand subjects were classified as stars in this project (see Section~\ref{sec:discussion}), only several tens were classified as artifacts. The vast majority of these were scattered light or saturation features, including the ejected material from the Cat's Eye Nebula (the hazy blue-green objects in Figure \ref{fig:top20-problem}), with none receiving high vote fractions for the other responses, indicating the image reduction pipeline is robust against such artifacts. Nevertheless, subjects like these, especially the identified population of stars, can be used to improve the automated image reduction pipeline to reduce their presence in future data sets, such as acting as a training set for deep learning models.

More examples of classified subjects can be found in Appendix~\ref{appendix:examples} for those classified as `Edge-On Disk', `Major Disturbance', `Clumps', and `Ring' (Figures \ref{fig:top20-edge-on} to \ref{fig:top20-ring}, respectively).
See Section~\ref{sec:discussion} for the number counts of subjects for each question and response; more detailed investigations of the classified objects, such as for identified `clumpy' galaxies, are left for future work.

\begin{table*}
	\centering
	\caption{57 gravitational lenses found in \ac{GZCD} and its corresponding Galaxy Zoo Mobile workflow after follow-up expert classification. Grades are scaled from 1.0 (least confident) to 3.0 (most probable lenses) as ranked by experts. The six lenses with names beginning with `EUCL' were first identified in the \textit{Euclid} survey. All others are newly identified by \ac{GZCD}.}
	\label{tab:lenses}
    \setlength{\tabcolsep}{6pt}
	\begin{tabular}{rrcccc}
		\hline
	    Name & H20 ID & grade & ra (deg) & dec (deg) & $m_{{\rm HSC}-i}$\\
        \hline
HSCJ180359.53+653913.3 & 17227 & 1.8 & 270.99803 & +65.653702 & 20.824 \\
HSCJ181322.08+664816.0 & 44151 & 3.0 & 273.34201 & +66.804446 & 19.759 \\
HSCJ181332.92+664830.6 & 44516 & 2.0 & 273.38719 & +66.808504 & 20.779 \\
HSCJ175324.78+661213.9 & 69795 & 1.4 & 268.35324 & +66.203873 & 19.946 \\
HSCJ175628.80+663851.6 & 85475 & 2.2 & 269.11998 & +66.647675 & 21.484 \\
HSCJ174516.59+664045.4 & 97816 & 3.0 & 266.31911 & +66.679290 & 20.184 \\
HSCJ180728.17+665805.6 & 117288 & 2.0 & 271.86736 & +66.968219 & 21.405 \\
HSCJ180129.51+673848.1 & 202458 & 1.0 & 270.37297 & +67.646685 & 21.375 \\
HSCJ175321.16+660848.3 & 214983 & 1.4 & 268.33816 & +66.146755 & 19.771 \\
HSCJ174613.93+662840.2 & 235798 & 3.0 & 266.55803 & +66.477826 & 20.121 \\
HSCJ175619.61+660944.8 & 273010 & 3.0 & 269.08169 & +66.162457 & 19.704 \\
HSCJ175422.02+672222.4 & 336139 & 1.8 & 268.59174 & +67.372883 & 19.949 \\
HSCJ180621.71+663023.3 & 495262 & 3.0 & 271.59048 & +66.506468 & 18.052 \\
HSCJ180436.99+662706.1 & 504385 & 3.0 & 271.15414 & +66.451682 & 18.094 \\
HSCJ175049.90+665454.5 & 509338 & 3.0 & 267.70790 & +66.915137 & 20.421 \\
HSCJ174951.93+665526.4 & 510408 & 1.2 & 267.46640 & +66.923999 & 20.175 \\
HSCJ174945.19+664636.6 & 511779 & 1.2 & 267.43830 & +66.776839 & 20.798 \\
HSCJ174949.03+661347.5 & 545792 & 3.0 & 267.45428 & +66.229861 & 18.858 \\
HSCJ181245.89+670241.2 & 597672 & 1.6 & 273.19121 & +67.044787 & 18.706 \\
HSCJ181058.39+664322.1 & 721210 & 1.2 & 272.74329 & +66.722808 & 17.287 \\
HSCJ181047.64+663555.5 & 724479 & 3.0 & 272.69849 & +66.598754 & 21.251 \\
HSCJ180152.17+660212.0 & 822910 & 1.0 & 270.46737 & +66.036658 & 18.858 \\
HSCJ181216.92+662144.5 & 876372 & 2.4 & 273.07052 & +66.362349 & 21.045 \\
HSCJ181125.34+662545.0 & 964551 & 1.0 & 272.85559 & +66.429170 & 20.095 \\
HSCJ180237.07+655052.2 & 973920 & 1.2 & 270.65447 & +65.847839 & 20.527 \\
HSCJ180008.01+663311.4 & 1019566 & 1.6 & 270.03339 & +66.553169 & 21.492 \\
HSCJ180443.36+654306.8 & 1047571 & 1.0 & 271.18065 & +65.718566 & 19.960 \\
HSCJ174702.33+661307.6 & 1080413 & 3.0 & 266.75972 & +66.218764 & 19.956 \\
EUCLJ175255.67+672542.9	& 1121829 & 3.0 & 268.23197 & +67.428585 & 20.373 \\
HSCJ175322.89+663141.1 & 1227621 & 1.4 & 268.34539 & +66.528092 & 20.945 \\
HSCJ180746.49+664843.8 & 1248193 & 2.2 & 271.94373 & +66.812153 & 21.433 \\
HSCJ180132.11+652551.4 & 1292299 & 2.4 & 270.38381 & +65.430937 & 21.497 \\
EUCLJ180216.86+652534.4 & 1292941 & 3.0 & 270.57026 & +65.426226 & 19.571 \\
HSCJ180147.29+652629.8 & 1293475 & 3.0 & 270.44704 & +65.441605 & 21.492 \\
HSCJ180927.19+664237.5 & 1303221 & 1.8 & 272.36327 & +66.710415 & 20.679 \\
HSCJ181006.49+663418.1 & 1305566 & 1.4 & 272.52703 & +66.571693 & 20.684 \\
HSCJ181344.70+663524.8 & 1342320 & 1.6 & 273.43626 & +66.590219 & 21.242 \\
EUCLJ175131.72+665425.2 & 1364482 & 3.0 & 267.88221 & +66.907003 & 18.846 \\
EUCLJ175804.74+661103.9 & 1387956 & 3.0 & 269.51977 & +66.184411 & 21.458 \\
HSCJ180611.55+665923.4 & 1439592 & 1.0 & 271.54811 & +66.989825 & 20.373 \\
HSCJ175209.70+670525.7 & 1469769 & 1.4 & 268.04041 & +67.090461 & 20.995 \\
HSCJ175749.19+674851.2 & 1514765 & 3.0 & 269.45494 & +67.814236 & 20.963 \\
EUCLJ180103.57+662743.2 & 1555317 & 3.0 & 270.26492 & +66.461998 & -- \\
HSCJ175205.80+654024.6 & 1616446 & 2.2 & 268.02418 & +65.673505 & 20.493 \\
HSCJ175910.99+663943.7 & 1633099 & 1.0 & 269.79579 & +66.662150 & 20.471 \\
HSCJ175700.77+665007.0 & 1788681 & 2.2 & 269.25320 & +66.835276 & 20.242 \\
HSCJ175702.03+665814.7 & 1809137 & 1.4 & 269.25845 & +66.970741 & 21.361 \\
HSCJ180215.58+652938.9 & 1825141 & 1.2 & 270.56490 & +65.494136 & 20.028 \\
HSCJ180151.75+653101.8 & 1826805 & 1.6 & 270.46564 & +65.517172 & 18.669 \\
HSCJ175909.19+652448.9 & 1833590 & 1.6 & 269.78829 & +65.413572 & 19.571 \\
HSCJ175957.54+652552.0 & 1835456 & 3.0 & 269.98974 & +65.431113 & 18.173 \\
EUCLJ180429.32+665508.1 & 1886777 & 3.0 & 271.12206 & +66.918792 & 19.419 \\
HSCJ180238.94+672735.9 & 1921420 & 1.6 & 270.66223 & +67.459962 & 19.288 \\
HSCJ180237.84+6726010.0 & 1931422 & 1.6 & 270.65766 & +67.436106 & 21.092 \\
HSCJ180017.57+661219.5 & 2016056 & 2.2 & 270.07323 & +66.205429 & 20.134 \\
HSCJ175936.12+661233.1 & 2048535 & 3.0 & 269.90050 & +66.209208 & 20.626 \\
HSCJ175809.55+664134.2 & 2063518 & 3.0 & 269.53978 & +66.692824 & 18.846 \\
        \hline
	\end{tabular}
\end{table*}

\begin{figure*}
	\centering
	\includegraphics[width=0.95\textwidth]{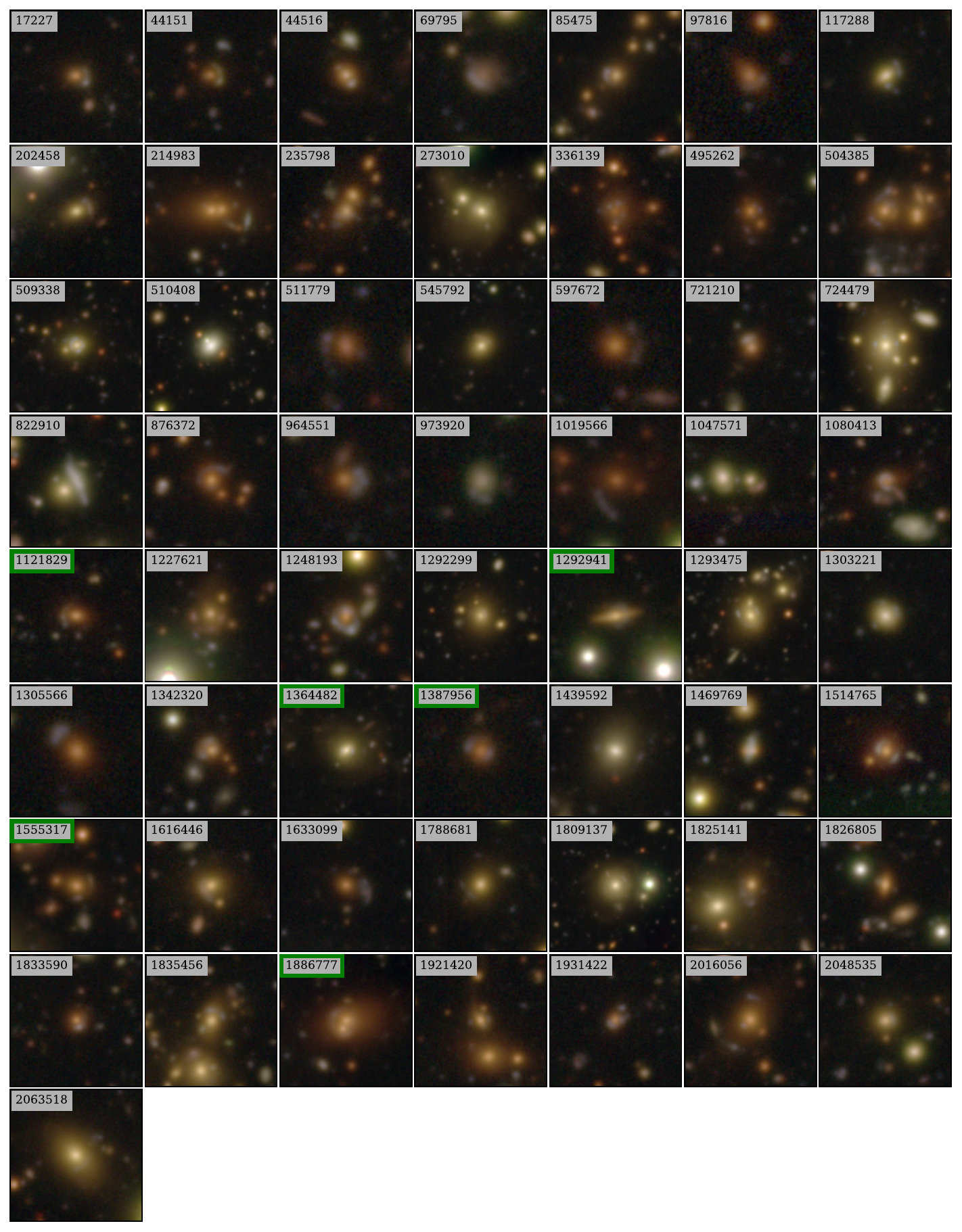}
    \caption{57 gravitational lenses found in \ac{GZCD} and its corresponding Galaxy Zoo Mobile workflow after follow-up expert classification. Subjects are listed in order of their H20 ID in accordance with Table \ref{tab:lenses}, with the majority being newly identified lenses in \ac{GZCD}. Green borders highlight those first identified by the \textit{Euclid} survey.}
    \label{fig:lenses}
\end{figure*}

\subsubsection{Tags, EROs and gravitational lenses}
\label{subsec:talk-tags}

In addition to aggregating volunteer votes for the decision tree of questions, another source of classifications comes from the tags that volunteers have assigned subjects on the Galaxy Zoo Talk boards - see Section~\ref{subsec:red-objects}. In that section it was highlighted that EROs should be tagged appropriately: across the objects below $z_{\rm phot}<2.5$, 2484 were tagged as `\#bright-red' or `\#extremely-red'; 789 tagged as `\#dim-red' or `\#dark-red'; 572 tagged as `\#bright-red-companion'; and 445 tagged as `\#dim-red-companion'. However, these samples were found to be highly impure due to the subjective nature of defining colours, with many having a more orange appearance akin to many of the objects in this data set. Hence, alternative algorithm-based methods may prove to be more effective in this regard, although the `\#bright-red-companion' tags provided a slightly purer sample: examples of which are shown in Figure \ref{fig:tagged-bright-red-companion} of Appendix \ref{appendix:examples}, which contain extremely red point-source objects away from the cutout centre. Elsewhere, a variety of tags were used to highlight objects volunteers found particularly beautiful or interesting, which were typically spiral, irregular or disturbed galaxies, as well as other rare objects - see Figure \ref{fig:tagged-interesting-beautiful} in Appendix \ref{appendix:examples} for some examples.

As discussed in Sections \ref{subsec:decision-tree} and \ref{subsec:lenses}, we also aimed to search this data set for gravitational lenses, including through a dedicated Galaxy Zoo Mobile workflow set up to find lenses in subjects retired early by Zoobot. Classifications from the main Galaxy Zoo workflow were searched with a selection cut of at least 6 votes for `Lens or arc', which from visual inspection led to a purer and equally complete sample compared to using cuts based on vote fractions. These 179 potential candidates, and all potential candidates from the Talk boards (467) and Galaxy Zoo Mobile (501), went through initial visual inspection vetting by the lead author to rule out clear false positives.
Across the entire project, 122 unique lens candidates were found as a result, with some duplicates obtained by different means: 32 from the main Galaxy Zoo workflow; 56 tagged by volunteers in the Talk boards, plus another 5 not tagged as lenses, but as `Rings'; and 63 from Galaxy Zoo Mobile. These were then classified by a team of experts in the Euclid Consortium's Strong Lensing Science Working Group using both HSC and \textit{Euclid} imaging \citep{2025arXiv250315324E}, narrowing the final list of lenses to the following 57: 15 new grade A lenses, 9 grade B lenses, 27 grade C lenses, and another 6 lenses already known to \textit{Euclid} (all grade A). The 57 lenses are shown in Table \ref{tab:lenses} and Figure \ref{fig:lenses}, and are listed in the main catalogue of the above publication.
Of these, 23 (2), 34 (11), and 23 (21) lenses were found (uniquely) through the main workflow, Talk tags, and Galaxy Zoo Mobile, respectively. 23 (40 per cent) were found by more than one method, with the vast majority of overlap coming from the main workflow and Talk tags (the latter finding all 23). Hence, most of the lenses were found through Talk tags, which proved to be more complete than the main workflow albeit with a higher rate of false positives, while the dedicated Galaxy Zoo Mobile workflow found by far the most candidates not seen by other methods. This highlights the importance of using multiple approaches to find rare objects where completeness is of great importance.

\subsection{Zoobot classifications}
\label{subsec:zoobot-classifications}

Once the project had been completed, a final run of Zoobot was performed on the subjects. Unlike the active learning cycle, this Zoobot model was reworked to use a new \textsc{EfficientNetV2\_S} architecture \citep{2021arXiv210400298T}. Once again, it was trained on colour images from all major Galaxy Zoo campaigns (GZ2, DECaLS/DESI, Hubble, Candels, Illustris, and now Cosmic Dawn itself), and predicted volunteer responses and associated 90 per cent upper and lower confidence intervals for question responses from all campaigns, though here we focus only on those relating to \ac{GZCD}. As with the active learning, the `rare features' question was again excluded from training and testing as it uniquely allowed multiple responses to be selected.

\begin{figure}
	\includegraphics[width=\columnwidth]{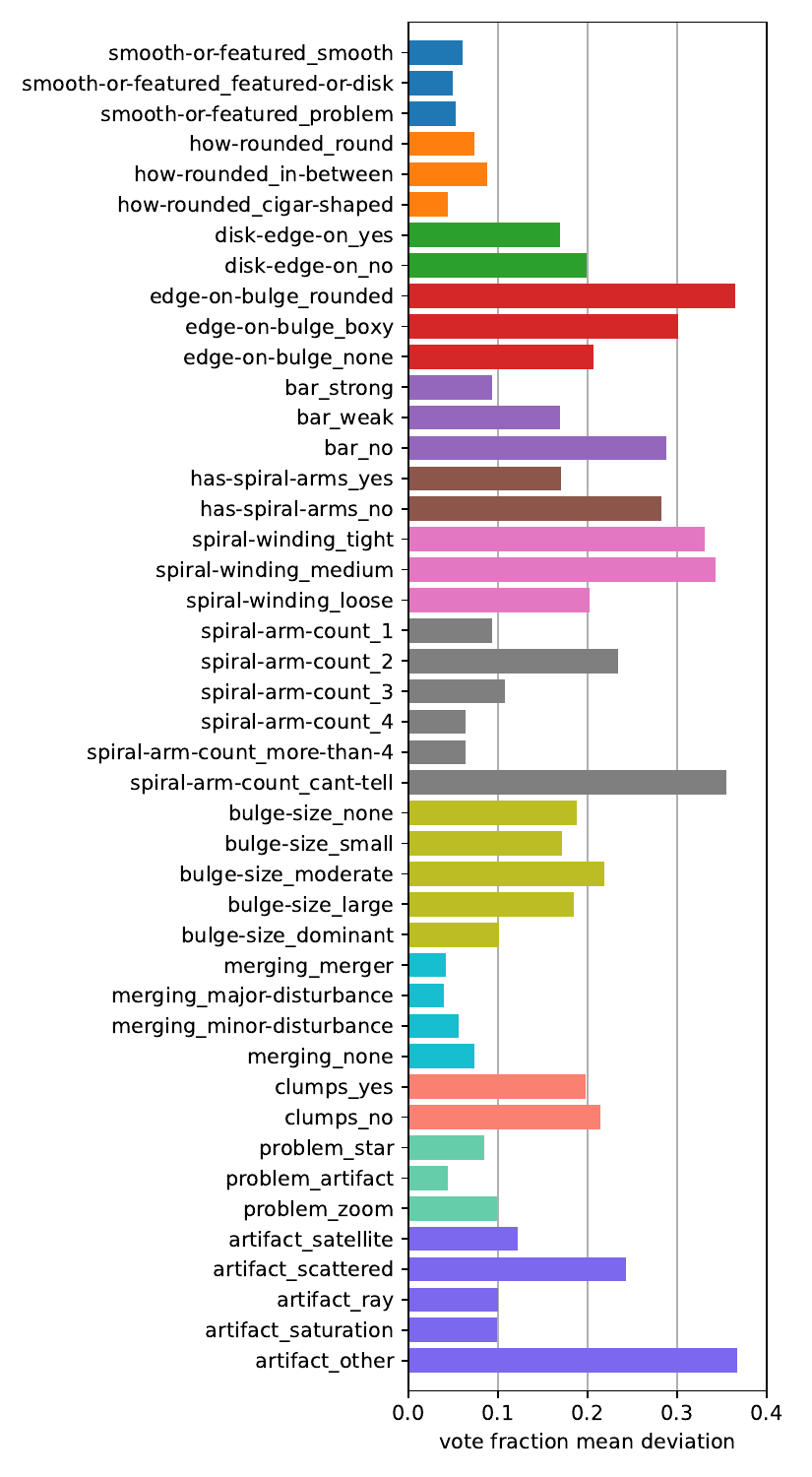}
    \caption{Mean absolute deviations of the vote fractions between Zoobot and volunteers for each possible response in the decision tree, limited to subjects with at least 40 volunteer responses. Responses are grouped by colour according to each question, and ordered approximately according to their level in the decision tree, with the new `problem' and `artifact' questions' responses listed at the end; `rare features' responses were not predicted by Zoobot so are not included. Out of 44 possible response options, the number of them with mean deviations below 0.1, 0.2 and 0.3 were 19, 30, and 38, respectively.}
    \label{fig:vote-frac}
\end{figure}

\begin{figure*}
	\includegraphics[width=0.8\textwidth]{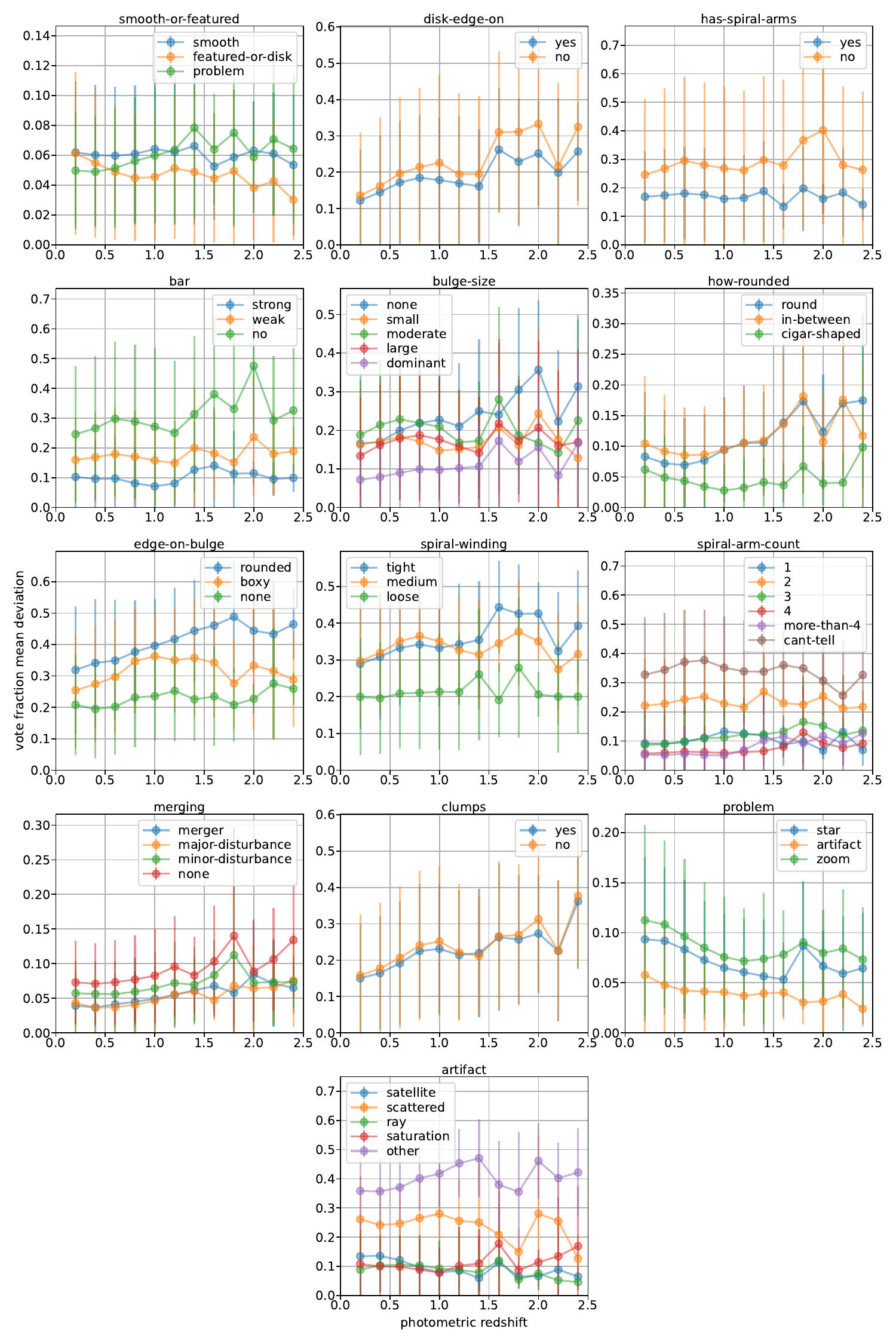}
    \caption{Mean absolute deviations of the vote fractions between Zoobot and volunteers as functions of photometric redshift, split between each question in the decision tree. Results are binned according to redshift with bins of width 0.2, and errorbars depict the standard deviation for each binned mean value. Plots are limited to subjects with photometric redshifts and at least 40 volunteer responses, and later bins contain far fewer subjects as per Figure \ref{fig:z_dist}.}
    \label{fig:vote-frac-vs-z}
\end{figure*}

\begin{figure*}
	\includegraphics[width=0.8\textwidth]{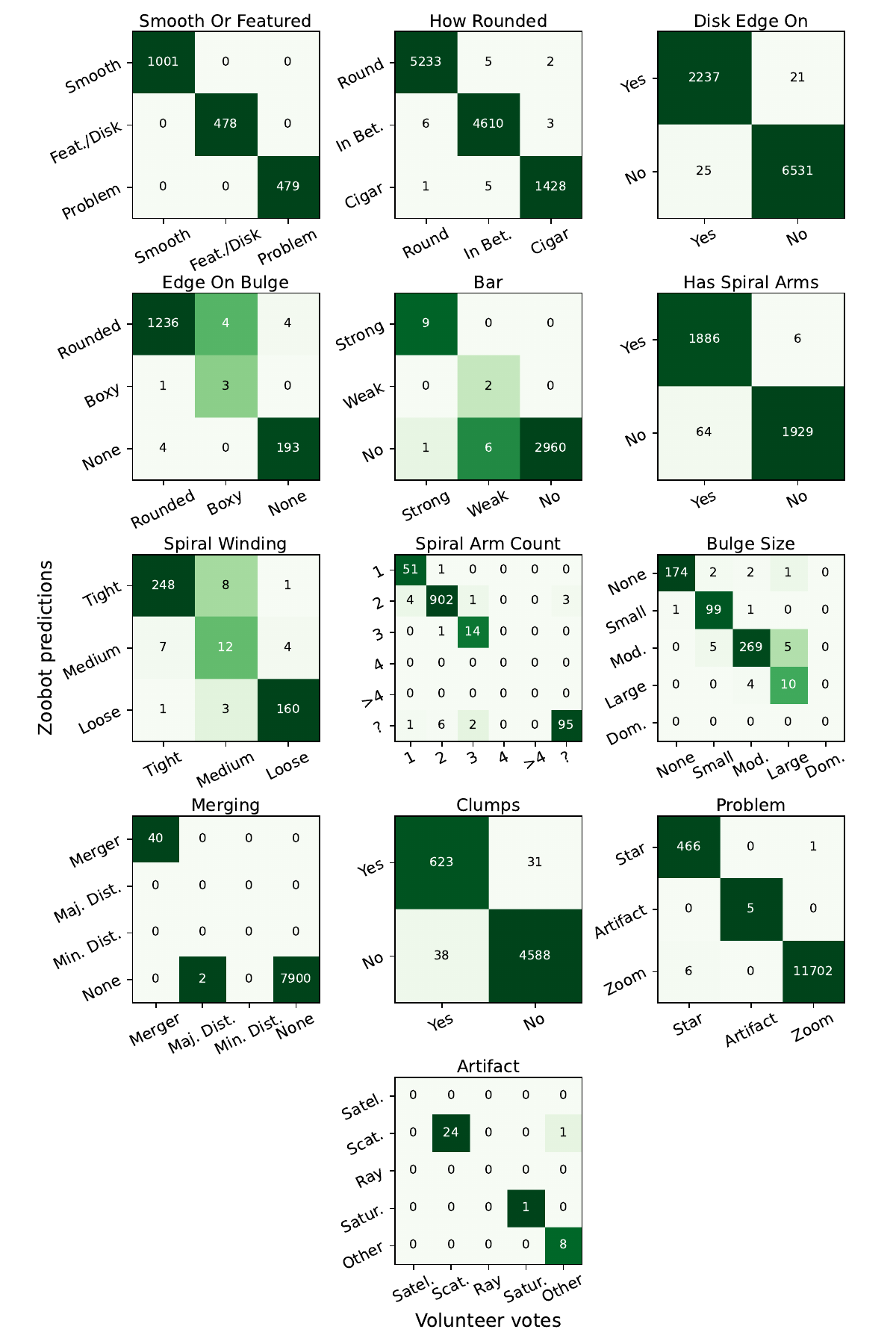}
    \caption{Confusion matrices of classifications from Zoobot and volunteers for each question in the decision tree. Matrices are limited to subjects with at least 40 volunteer votes, and further limited to those with `high confidence' classifications (i.e. those with volunteer vote fractions >0.8) and, for a given question, at least 5 volunteer votes to reduce the `ground truth' uncertainty.}
    \label{fig:confusion-matrix}
\end{figure*}

Figure \ref{fig:vote-frac} shows the mean absolute deviations between Zoobot's predicted vote fractions and those of volunteers. Across the possible response options, 43 per cent, 68 per cent, and 86 per cent had mean deviations below 0.1, 0.2 and 0.3, respectively. These deviations are typically larger than the $\sim$10 per cent seen in previous iterations \citep{2022MNRAS.509.3966W,2025arXiv250315310E} due to smaller sources from deeper imaging and increased confusion noise from background galaxies. Some of the more accurately predicted responses are for the initial `smooth-or-featured' question, the `how-rounded' question for `Smooth'-voted galaxies, and the `merging' question, which received the largest number of volunteer votes due to the structure of the decision tree (Figure \ref{fig:decision-tree}) combined with the majority of subjects being classed as `Smooth' in this data set (see Figure \ref{fig:morphology-vs-z}). Also, despite the `problem' and `artifact' questions being new additions to this Galaxy Zoo iteration, most of their responses were still predicted within 10 per cent of the volunteer's results. The largest deviations came from perhaps less clear-cut and more subjective response options, such as the shapes of edge-on bulges, the tightness of spiral arm winding, the `Can't Tell' option for spiral arm count and `Other' for artifact type: volunteers' responses for these can vary more widely while Zoobot may choose more specific responses based on hard-to-spot features.

To analyse these differences further, the mean absolute deviations are shown again in Figure \ref{fig:vote-frac-vs-z}, now binned as functions of redshift. As can be seen, there are large uncertainties in the binned deviations between Zoobot and volunteer vote fractions, however the deviations do increase or decrease with redshift for several question responses, such as the increase for all `disk-edge-on' and `clumps' responses and the decrease for all `problem' responses. Interestingly, for the majority of responses with the highest deviations in Figure \ref{fig:vote-frac}, there is very little dependence on redshift, suggesting these deviations are affected primarily by limited training of Zoobot on this survey data, rather than an unavoidable limitation of the model (or volunteer accuracy) when applied to higher-redshift objects.

It should be noted that the branching decision tree leads to some questions receiving far fewer total votes than others for a given subject (such as an obvious artifact receiving little to no votes for questions along the `Features or Disk' branch). Hence, for each decision tree question, we can further limit the data set to only subjects that have received at least 5 total votes in response to that question, to avoid conflating errors in Zoobot predictions with any uncertainty in the `ground truth' of volunteer votes.
This additional restriction served to greatly reduce errors, reducing the mean deviations in Figure \ref{fig:vote-frac} to around 0.12 or lower for nearly all responses. Hence, it is clear that, for a given subject, Zoobot struggles at accurately predicting volunteer vote fractions for questions which the volunteers are not providing sufficient votes for (for which the vote fractions are, and should be predicted as, zero for all responses) -- this is in reality an edge case and Zoobot otherwise excels at predicting volunteer vote fractions.

Confusion matrices for each question are presented in Figure \ref{fig:confusion-matrix}, comparing the numbers of subjects classified as each response by volunteers and by Zoobot. Corresponding classification metrics are given in Table \ref{tab:confusion_matrix_metrics}, and as with the other comparisons in this section, these matrices and metrics are limited to subjects with at least 40 volunteer responses (i.e. `fully classified' by volunteers). In accordance with the above, we only consider subjects that have received at least 5 total votes in response to a given question, and here they are further limited to those with `high confidence' classifications (defined as those with volunteer vote fractions >0.8) as presented in \cite{2025arXiv250315310E}: Figure 13 of that work highlights that Zoobot prediction errors tend to occur primarily when a subject's volunteer vote fraction lies within the range 0.4-0.6, with Zoobot becoming increasingly accurate for those with a more confident volunteer vote fraction (closer to 0 or 1). Even so, for the \ac{GZCD} data set, excluding subjects with `uncertain' labels produced only a minor effect on Zoobot accuracy.
It is clear from the results that Zoobot performed exceptionally across the majority of questions and responses, even predicting with 100 per cent accuracy how volunteers would vote for responses to the first question (`smooth-or-featured'). It should be noted however that this is fairly expected behaviour for a well-trained model, as it is testing here on the same data set used to train it (albeit only one subset of the training data). The model primarily struggled on accurately predicting volunteer votes for `Boxy' edge-on bulges, `Weak' bars, `Medium' spiral winding, `Large' bulge sizes, and those undergoing `Major Disturbances', however it is clear from Figure \ref{fig:confusion-matrix} that in this subject set the ground truth sets for these responses were very small.

\begin{figure}
	\includegraphics[width=\columnwidth]{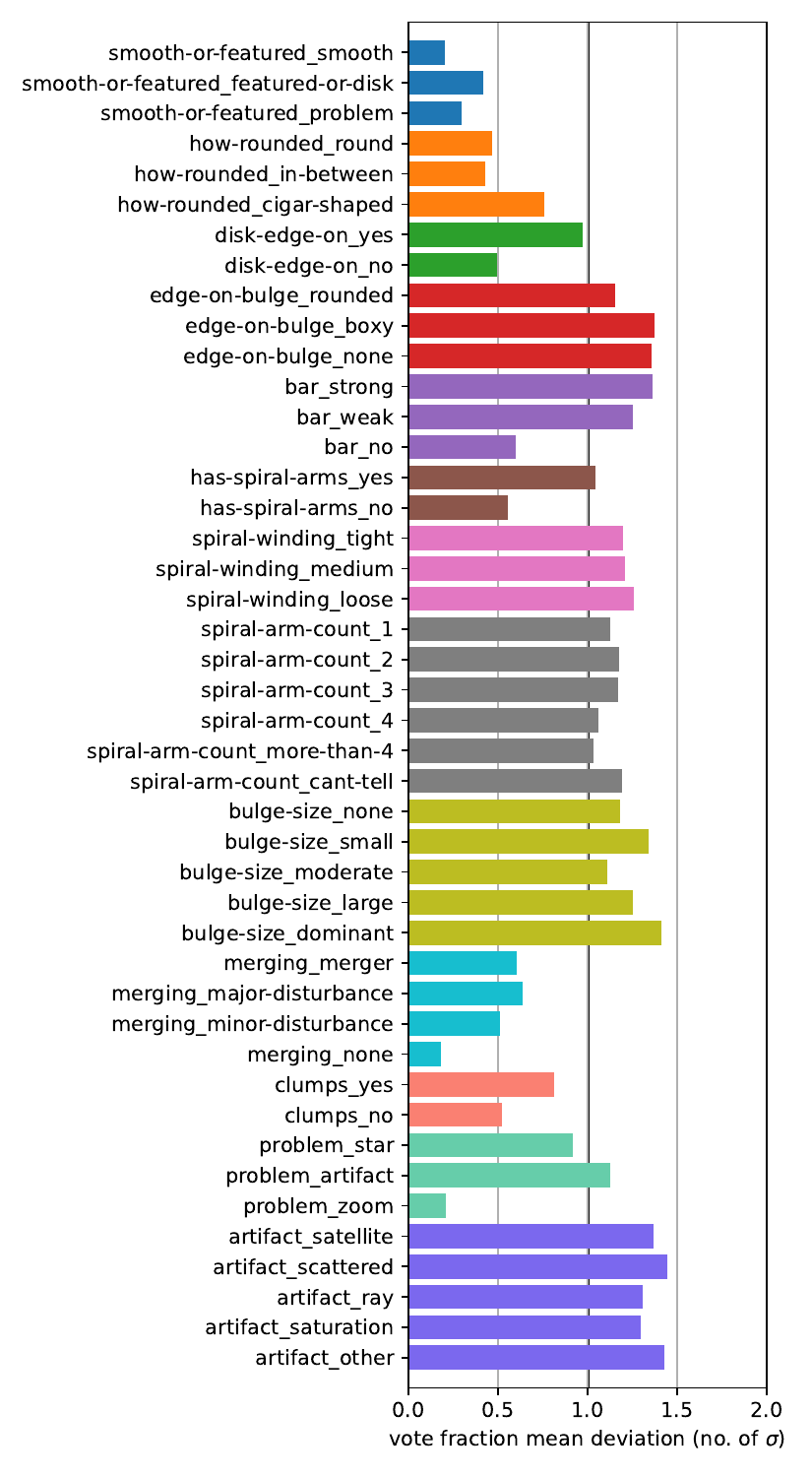}
    \caption{Mean absolute deviations of the vote fractions between Zoobot and volunteers for each possible response in the decision tree, limited to subjects with at least 40 volunteer responses, given in terms of the number of $\sigma$ based on the 95 per cent confidence intervals predicted by Zoobot. The format follows the same as Figure \ref{fig:vote-frac}.}
    \label{fig:vote-frac-sigmas}
\end{figure}

In summary, this Zoobot model agrees very well with the volunteers for responses with many votes and subjects with a high level of confidence (e.g. vote fractions >0.8), though can somewhat struggle on more subjective response options and, for some questions like `disk-edge-on' and `clumps', subjects with high photometric redshifts. That said, Zoobot also predicts uncertainties (95 per cent confidence intervals) on its vote fractions: Figure \ref{fig:vote-frac-sigmas} shows the mean average vote fraction deviations once each individual deviation is first converted to a number of $\sigma$ when taking the 95 per cent confidence intervals to be 1.645$\sigma$. As shown, the uncertainties that Zoobot predicts are, on average, appropriately sized (with a mean average absolute deviation of 0.95$\pm$0.39$\sigma$ across all responses), despite some clearly being over-predicted while others under-predicted, with the uncertainties on the first question especially over-predicted (i.e. Zoobot is underconfident at predicting smooth vs. featured galaxies). It is worth noting the possibility that Zoobot may be picking up on subtle features missed by volunteers -- which would make its predictions more accurate than volunteers -- however, the results shown here are for subjects classified by at least 40 volunteers and hence this scenario is highly unlikely. A more likely explanation is that Zoobot simply requires a larger training set with similar deep ground-based imaging of subjects that correspond to responses for which the current sample size is comparatively small. This will allow the model to match or exceed the otherwise excellent performance it has demonstrated for all other responses, both in GZCD and previous Galaxy Zoo iterations.

\subsection{Quantitative Tests of our Qualitative Classifications}
\label{subsec:quantitative-test}

To test the reliability of the classifications made by volunteers and Zoobot, we compared these to quantitative fits applied to HSC-$r$ and -$g$ band cutout images of almost 3000 subjects with HSC-$i$ band apparent magnitudes brighter than 21. The \textsc{GALFIT} software \citep{2010AJ....139.2097P} was used to obtain fits to their optical surface brightness distributions, assuming that they follow a S\'ersic profile \citep{1963BAAA....6...41S,1968adga.book.....S} in which we allow the S\'ersic index $n$ to vary freely between 1.0 (appropriate for a pure disk galaxy with no bulge) and 4.0 (appropriate for a pure spheroid -- bulge or elliptical -- with no disk present). The {\it quantitative} measures of the surface brightness distribution from these fits were then compared with the {\it qualitative} classifications of the Galaxy Zoo volunteers for this NEP galaxy subsample. 

\begin{table}
	\centering
	\caption{Classification metrics for Zoobot-predicted vote fractions as compared to those of volunteers, for subjects with at least 40 volunteer votes and high confidence (>0.8) volunteer vote fractions, and, for a given question, at least 5 volunteer votes to reduce the `ground truth' uncertainty. Metrics use `macro' averaging: for `micro' averaging the accuracy is unchanged and all other metrics equal the accuracy values.}
	\label{tab:confusion_matrix_metrics}
    \setlength{\tabcolsep}{5pt}
	\begin{tabular}{cccccc}
		\hline
		Question & Count & Accuracy & Precision & Recall & F1\\
        \hline
        Smooth Or Featured & 1958 & 1.00 & 1.00 & 1.00 & 1.00\\
        How Rounded & 11293 & 1.00 & 1.00 & 1.00 & 1.00\\
        Disk Edge On & 8814 & 0.99 & 0.99 & 0.99 & 0.99\\
        Edge On Bulge & 1445 & 0.99 & 0.91 & 0.80 & 0.84\\
        Bar & 2978 & 1.00 & 1.00 & 0.72 & 0.78\\
        Has Spiral Arms & 3885 & 0.98 & 0.98 & 0.98 & 0.98\\
        Spiral Winding & 444 & 0.95 & 0.82 & 0.82 & 0.82\\
        Spiral Arm Count & 1081 & 0.98 & 0.95 & 0.92 & 0.94\\
        Bulge Size & 573 & 0.96 & 0.91 & 0.88 & 0.89\\
        Merging & 7942 & 1.00 & 1.00 & 0.67 & 0.67\\
        Clumps & 5280 & 0.99 & 0.97 & 0.97 & 0.97\\
        Problem & 12180 & 1.00 & 1.00 & 1.00 & 1.00\\
        Artifact & 34 & 0.97 & 0.99 & 0.96 & 0.97\\
		\hline
	\end{tabular}
\end{table}

\textsc{GALFIT}'s quantitative analysis begins by fitting ellipses to the observed isophotes to determine the semi-major and semi-minor axes. Their ratio, $b/a$, is an isophotally-averaged measure of how elongated the galaxy appears. We first find that this simple fitted quantitative value corresponds very closely with the qualitative judgements made by the volunteers. Concentrating on galaxies with good consensus among the volunteers (vote fractions $f$>0.7) for the `How rounded is it?' question, we find that volunteer-voted `Round', `In-Between' and `Cigar-Shaped' galaxies have average fitted ($b/a$) ratios of 0.84, 0.57, and 0.24, respectively. Likewise, concentrating on galaxies with good volunteer consensus for the `Could this be a disk viewed edge-on?' question, and focusing on the galaxies best fitted by particularly low S\'ersic indices ($n$<1.5) suggestive of edge-on disks, those voted `Yes' had average fitted ($b/a$) ratios of 0.31, in strong contrast to 0.65 for those voted `No'.

Next, when splitting the fitted galaxies into two groups -- `disk-like' galaxies with $n$=1.00-2.00 versus `spheroid-like' with $n$=2.01-4.00 -- we find a clear distinction made by the volunteer classifiers for the `How rounded is it?' question: even based simply on highest-majority responses without further selection cuts (see Section~\ref{sec:discussion}), the roundedness classification frequencies of the disk-like galaxies for `Round', `In-Between' and `Cigar-shaped' were found to be 0.28, 0.52 and 0.20 respectively, in strong contrast to those for the spheroid-like galaxies -- 0.52, 0.40 and 0.08. These same strong differences were also measured by the Zoobot automated classifier: 0.33, 0.55 and 0.18 for disk-like galaxies as opposed to 0.57, 0.40 and 0.03 for the spheroid-like galaxies.
An additional test was done after restricting the sample to only those galaxies classified by the volunteers as `Smooth' (vote fraction for the `smooth or featured' question of $f_{\rm smooth}>0.7$; see Section~\ref{sec:discussion}) with strong agreement on their shapes ($f_{\rm completely-round}>0.8$). These selection cuts reduce the morphological sample size (442 subjects), but increase its quality and purity by focusing on subjects with high-confidence classifications. The results support those of the larger full sample: the roundedness classification frequencies of disk-like galaxies for `Round', `In-Between' and `Cigar-shaped' were found to be 0.31, 0.57 and 0.12, in contrast to 0.74, 0.26 and 0.00 for spheroid-dominated galaxies.
Since most categories contain hundreds of galaxies, the typical uncertainties in these mean values are on the order of 0.01. Thus, the classification differences for quantitatively disk-like and spheroid-like galaxies are overwhelmingly significant. These classification differences are exactly what one would expect, since galaxies with a dominant disk are intrinsically flat, and so they should be much less likely to be viewed as `round' and more likely to be viewed as highly elongated compared with the more intrinsically round spheroid-dominated galaxies.

One further test is a comparison of the low-$n$ (disk dominated) and high-$n$ (spheroid dominated) galaxies to their Galaxy Zoo classifications regarding spiral arms (i.e. `Is there any sign of a spiral arm pattern?'). Again, even based simply on the highest-majority responses without further selection cuts, 46 per cent of the galaxies with S\'ersic indices $n$=1.00-2.00 are voted as having spiral arms, compared with only 32 per cent of those with $n$=2.01-4.00. 
Additionally, we find similar results when starting from the volunteer classifications: the average fitted S\'ersic index of galaxies confidently voted as having spiral arms ($f_{\rm yes}>0.7$) is 1.29, as opposed to 1.60 for those confidently voted as not having spiral arms ($f_{\rm no}>0.7$).
These differences, which are also highly significant, provide another independent quantitative indication that the Galaxy Zoo volunteers are more likely to identify spiral arms in the later-type galaxies based on objective numerical S\'ersic surface brightness fitting.

Finally, although our viewing angle usually does not provide an edge-on view of a disk galaxy, a significant minority fraction of the galaxies are classified by volunteers as having an edge-on disk. Again, this matches with our S\'ersic surface brightness fits: the average fraction of galaxies for which they said `Yes' to `Could this be a disk viewed edge-on?' is significantly higher (28 per cent) in the disk-dominated galaxies than it is for the spheroid-dominated galaxies (21 per cent, with an uncertainty smaller than 1 per cent).
A similar difference persists when using selection cuts to restrict the sample to high-confidence classifications ($f_{\rm featured}>0.3$ and $f_{\rm yes-edge-on-disk}$>0.8). Among that subset of 1061 S\'ersic-fitted galaxies, 20$\pm$1.4 per cent of disk-like galaxies were voted with high confidence as having edge-on disks, compared with only 14$\pm$3 per cent of spheroid-dominated galaxies.

All of these simple quantitative comparisons serve as `confidence builders' for our galaxy morphology classifications, since they are indeed supported by purely objective measurements of the surface brightness profiles of the same galaxies.

\section{Data Release}
\label{sec:discussion}

Alongside this paper, we release catalogues containing the aggregated classifications made by volunteers and by Zoobot for \ac{GZCD}, along with the set of PNG/JPG cutout images used in the project and the associated metadata for each subject, all available from \href{https://doi.org/10.5281/zenodo.17200992}{Zenodo} \citep{pearson_2025_17200992}. We present an overview here, and additional documentation can be found alongside the data itself. All subjects (both cutouts and tabular data) are identified by their H20 IDs, and, in addition to the data mentioned in Section~\ref{subsec:metadata}, the metadata includes the Galaxy Zoo Talk board tags assigned to each subject as well as API links to various search databases. As discussed in Section~\ref{subsec:aggregated-classifications}, the data set consists of 45\,742 subjects with either photometric redshifts $z_{\rm phot}<2.5$ or for which the photometric redshift estimation failed.
The cutouts are split into two groups: the 15\,745 PNGs fully classified by volunteers prior to Zoobot active learning, and the 29\,997 JPGs classified with the help of active learning. Hence, many of those in the latter set will have less than 40 votes recorded in the classification catalogues.

The volunteer catalogue contains the number of votes for each question's responses for a given subject, the corresponding vote fractions $f$ (the number of votes for a response divided by the number of volunteers answering that question), as well as the total number of volunteers who classified it. The Zoobot catalogue contains its predicted vote fractions, along with upper and lower 90 per cent confidence bounds on those values, and the predicted fraction of volunteers answering a given question, although the `rare features' question is excluded for Zoobot as it was not trained on that question. Different versions of the classification catalogues are provided, with headers in the format of `question\_answer' or as a multi-index header for a more easily-readable format. There are also alternative versions designed for ease of use that contain only the `leaf fractions' - the vote fractions for the highest-predicted answers to each question - by setting any predicted vote fraction to `NaN' if the vote fractions from all preceding questions leading to that question multiply to 0.5 or less, as a rule-of-thumb for `is this question relevant to this image' (such as avoiding answering `What type of artifact is it?' for subjects mostly classified as `Smooth' or `Features or Disk').

\begin{table}
	\centering
	\caption{Example suggested selection cuts exploring the aggregated vote fractions $f$, and the number counts $N_{\rm sample}$ of subjects meeting those cuts for classifications made by volunteers and by Zoobot. Number counts for volunteer classifications are limited to subjects with at least 10 volunteer votes. See Table 2 of \protect\cite{2025arXiv250315310E} for more suggested cuts, which broadly apply to this data set as well. The table here includes suggested cuts for questions and responses not included in such previous works, with $f_{\rm problem}$ representing vote fractions for the `Star, Artifact, or Bad Zoom' response. For `Lens or Arc', `Irregular' and `Ring', no number counts are given for Zoobot as it was not trained to predict responses to the `rare features' question.}
	\label{tab:selection-cuts}
    \setlength{\tabcolsep}{5pt}
	\begin{tabular}{llcc}
		\hline
		Sample & Selection Approximate Cut & $N_{\rm sample}^{\rm volunteers}$ & $N_{\rm sample}^{\rm Zoobot}$\\
        \hline
        Smooth & $f_{\rm smooth}>0.7$ & 5943 & 13912\\
        Features or Disk & $f_{\rm featured}>0.3$ & 6328 & 6271\\
        Star & $f_{\rm problem}>0.3$ & 1333 & 3664\\
             & $f_{\rm star}>0.6$ &  & \\
        Artifact & $f_{\rm problem}>0.3$ & 52 & 79\\
             & $f_{\rm artifact}>0.5$ &  & \\
        Bad Image Zoom & $f_{\rm problem}>0.6$ & 1149 & 1551\\
             & $f_{\rm bad\ image\ zoom}>0.6$ &  & \\
        Edge-On Disk & $f_{\rm featured}>0.3$ & 1381 & 1312\\
             & $f_{\rm yes\ edge-on}>0.8$ &  & \\
        Any Bar & $f_{\rm featured}>0.3$ & 243 & 72\\
             & $f_{\rm not\ edge-on}>0.6$ &  & \\
             & $f_{\rm no\ bar}<0.2$ &  & \\
        Spiral & $f_{\rm featured}>0.3$ & 2485 & 2348\\
             & $f_{\rm not\ edge-on}>0.6$ &  & \\
             & $f_{\rm yes\ spiral\ arm\ pattern}>0.6$ &  & \\
        Clumps & $f_{\rm featured}>0.3$ & 691 & 662\\
             & $f_{\rm yes\ clumps}>0.7$ &  & \\
        Lens or Arc & $f_{\rm problem}<0.3$ & 14 & -\\
             & $f_{\rm lens\ or\ arc}>0.3$ &  & \\
        Irregular & $f_{\rm problem}<0.3$ & 360 & -\\
             & $f_{\rm irregular}>0.2$ &  & \\
        Ring & $f_{\rm problem}<0.3$ & 75 & -\\
             & $f_{\rm ring}>0.5$ &  & \\
		\hline
	\end{tabular}
\end{table}

Regardless of the catalogue, these vote fractions can be used to place selection cuts so as to interrogate the data set. For example, one may use $f_{\rm smooth} > 0.7$ (where $f_{\rm smooth}$ corresponds to the header `smooth-or-featured\_smooth') to select a fairly pure sample of galaxies classed as `Smooth'. Such selection cuts can also be combined in order to correctly interpret classifications made further along the branches of the decision tree: one might extract edge-on disks by specifying $f_{\rm featured}>0.3$ and $f_{\rm yes\ edge-on}>0.8$ (regarding headers `smooth-or-featured\_featured-or-disk\_fraction' and `disk-edge-on\_yes', respectively). It should be noted that there is no perfect cut for all users, as the balance of purity and completeness will depend on your science goal: a strict cut produces a purer sample, say, for follow-up observations, while a more relaxed cut would produce a sample whose distribution can be weighted by the subjects' vote fractions. To help obtain fairly pure samples, some examples of approximate selection cuts are provided in Table \ref{tab:selection-cuts} based on visual inspection, and more can be found in \cite{2022MNRAS.509.3966W}. In particular, the table provides cuts for questions and responses not previously included in recent data releases, including `Artifact', `Clumps' and some `rare features' responses, although for all cuts the reader should take care to examine the purity of their samples and adjust the cuts accordingly. For each of these selection cuts, the table also provides the corresponding number counts of subjects as classified by volunteers and by Zoobot.

It is clear from the table that for the first question, many galaxies have not been confidently classified, that is, have not received high vote fractions for any of the three responses. This highlights that these votes are entirely based on visual morphology rather than true morphology, and hence limited by image resolution and confusion noise. As mentioned in Sections \ref{subsec:weightings-and-debiasing} and \ref{subsec:aggregated-classifications}, the vote fractions have not been debiased with regard to redshift and so the majority of objects with $z_{\rm phot}\gtrsim1.0$ are seeing-limited and hence likely classified as `Smooth'.
With the \ac{GZCD} data set spread across a fairly wide redshift range, it is also possible that many of the galaxies are indeed less clearly classifiable due to unusual features or in-progress morphological evolution.
It should be noted that, given the PSF and lack of clear diffraction spikes for all but the brightest of stars in the deep HSC imaging, one area of difficulty in this Galaxy Zoo iteration came from faint stars being selected as subjects in the cutout-creation process, which made it challenging for volunteers (and hence a retrained Zoobot) to distinguish between stars and smooth spherical galaxies. As a result, the catalogue includes a large number of objects classified as stars, which may well be PSF-limited stars or smooth galaxies matching their appearance: we caution the reader to take care when utilising the data set for such objects, and highlight that a stricter cut of $f_{\rm smooth}>0.8$ for example tends toward more elliptical-shaped galaxies.

\section{Conclusion}
\label{sec:conclusion}

In the era of Big Data in astronomy, citizen science provides a crucial step in mining vast quantities of imaging data. In this work, we present an overview and data release for Galaxy Zoo: Cosmic Dawn (GZCD), an iteration of the Galaxy Zoo project on the Zooniverse citizen science platform for the morphological classification of galaxies. The data release features catalogues of morphological classifications made by over 10\,000 volunteers, and by Galaxy Zoo's deep learning model Zoobot, for over 45\,000 subjects (including over 41\,000 cutouts with central galaxies), and is accompanied by associated metadata and the cutout images as shown to volunteers.

The subjects in this iteration came from the Hawaii Twenty Square Degree survey as part of the wider Cosmic Dawn survey, featuring deep multiband Hyper Suprime-Cam (HSC) imaging from six square degrees of the Euclid Deep Field North, down to a depth of $m_{{\rm HSC}-i} = 21.5$ for the purpose of resolving galaxy morphology. With an updated decision tree of questions, volunteers were asked to identify clumpy galaxies and tag rare objects like gravitational lenses and extremely red objects, which are of particular interest given the wide redshift range of this data set (up to $z_{\rm phot}=2.5$; median of 0.42).
The final morphology catalogues in this data release contain a wide range of classified objects within the Euclid Deep Field North (EDFN), with suggested selection cuts providing a means of interrogating the results for a given science case, such as accessing a pure sample of edge-on disks, spirals, clumpy galaxies, and rarer irregular or ring galaxies. Given the sky coverage, this rich data set provides an opportunity to extract objects in the EDFN for further examination with deep \textit{Euclid} imaging and high-resolution follow-up observations. For example, the data has already given rise to 51 newly discovered gravitational lenses after expert vetting of HSC and \textit{Euclid} imaging of lens candidates found by volunteers. Additionally, with the ground-based HSC's seeing comparable to that of the Vera C. Rubin Observatory, the classified images will act as a valuable truth set for training models to classify not only HSC imaging from the Ultraviolet Near-Infrared Optical Northern Survey (UNIONS), but also the wealth of data from the newly begun Rubin Legacy Survey of Space and Time (LSST).

The Galaxy Zoo project allows for the visual inspection of galaxies by humans, in order to correctly classify their apparent morphologies and for the serendipitous discovery of new and interesting features that would be missed by machine learning algorithms. On the other hand, such work requires many (typically up to 40) volunteers per subject image for the classifications to be rigorous, and can therefore take a long time for a data set to be fully classified. Galaxy Zoo's deep learning model, Zoobot, provides a two-stage method for circumventing this. First, Zoobot can separate out (retire) the easier subjects while leaving volunteers with more challenging and interesting objects, whose classifications are then used to improve (fine-tune) Zoobot in an active learning cycle. Second, a fully fine-tuned Zoobot can confidently predict how volunteers would classify each image in a fraction of the time. Here, we find that the active learning cycle aided in retiring $\sim$78 per cent of subjects prior to them reaching 40 volunteer classifications, with the majority requiring less than 10 before Zoobot was confident enough to retire them. Then, applying a fully fine-tuned Zoobot to the whole subject set after project completion resulted in 68 per cent of vote fractions having mean deviations from volunteers below 0.2. Although larger than those in previous iterations, suggesting that further training is required on similar deep surveys, Zoobot's predicted uncertainties were, on average, consistent with the discrepancies. As new iterations of Galaxy Zoo take place, the model will continue to improve and new volunteer classifications will help us explore galaxy evolution over ever more significant fractions of the age of the Universe.

\section*{Acknowledgements}

This publication has been made possible by the participation of more than 10\,000 volunteers in the Galaxy Zoo project. We also gratefully acknowledge the contributions to this work by the \textit{Euclid} Strong Lensing Working Group for their visual inspection efforts of the strong gravitational lens candidates, and we thank the anonymous referee for their useful comments that helped improve the paper.

\ac{GZCD} is partly supported by the ESCAPE project, which aims to bring together the astronomy, astroparticle and particle physics communities to support open science, according to FAIR (Findable, Accessible, Interoperable and Reusable) principles. ESCAPE - The European Science Cluster of Astronomy \& Particle Physics ESFRI Research Infrastructures has received funding from the European Union's Horizon 2020 research and innovation programme under the Grant Agreement no. 824064.
The Cosmic Dawn Center (DAWN) is funded by the Danish National Research Foundation under grant DNRF140. The Hyper Suprime-Cam (HSC) collaboration includes the astronomical communities of Japan and Taiwan, and Princeton University.

JP and SS were partly supported by the ESCAPE project. JP, HD, and SS are supported by the ACME, ELSA, and OSCARS projects. ”ACME: Astrophysics Centre for Multimessenger studies in Europe”, ”ELSA: Euclid Legacy Science Advanced analysis tools”, and ”OSCARS: Open Science Clusters’ Action for Research and Society” are funded by the European Union under grant agreement no. 101131928, 101135203, and 101129751, respectively; ELSA is also funded by Innovate UK grant 10093177. Views and opinions expressed are however those of the authors only and do not necessarily reflect those of the European Union.
MW is a Dunlap Fellow. The Dunlap Institute is funded through an endowment established by the David Dunlap family and the University of Toronto.
LFF was partially supported by the National Science Foundation (NSF) award IIS-2334033.
BDS acknowledges support through a UK Research and Innovation Future Leaders Fellowship [grant number MR/T044136/1] and its renewal [grant number MR/Z000076/1].
DBS gratefully acknowledges support from NSF Grant 2407752.
IS acknowledges NASA grant 80NSSC24K1489.
WJP has been supported by the Polish National Science Center project UMO-2023/51/D/ST9/00147. This research was conducted under the agreement on joint mobility projects for the years 2024-2025 between the Polish Academy of Sciences and the National Science and Technology Council in Taiwan.

This work made use of \textsc{Astropy}:\footnote{\tt \url{http://www.astropy.org}} a community-developed core Python package and an ecosystem of tools and resources for astronomy \citep{2013A&A...558A..33A,2018AJ....156..123A,2022ApJ...935..167A}. Data analysis made use of the Python packages \textsc{NumPy} \citep{2020Natur.585..357H} and \textsc{SciPy} \citep{2020SciPy-NMeth}. Some figures were made with the Python package \textsc{matplotlib} \citep{Hunter:2007}.

\section*{Data Availability}

The data released alongside this work, including volunteer and Zoobot classifications as well as subject images and metadata, can be accessed on Zenodo at \href{https://doi.org/10.5281/zenodo.17200992}{https://doi.org/10.5281/zenodo.17200992} \citep{pearson_2025_17200992}, where they are publicly available for download and use; see the Zenodo page for more details.


\bibliographystyle{mnras}
\bibliography{bibliography}



\appendix

\section{Example Classifications}
\label{appendix:examples}

Here we provide more examples of classified subjects with some of the highest volunteer vote fractions for their response options, as referred to in Section~\ref{subsec:aggregated-classifications}. Figure \ref{fig:tagged-interesting-beautiful} provides an array of objects tagged by volunteers with tags such as ``beautiful'', ``interesting'', or ``needs-more-research''.

\begin{figure}
	\centering
	\includegraphics[width=0.9\columnwidth]{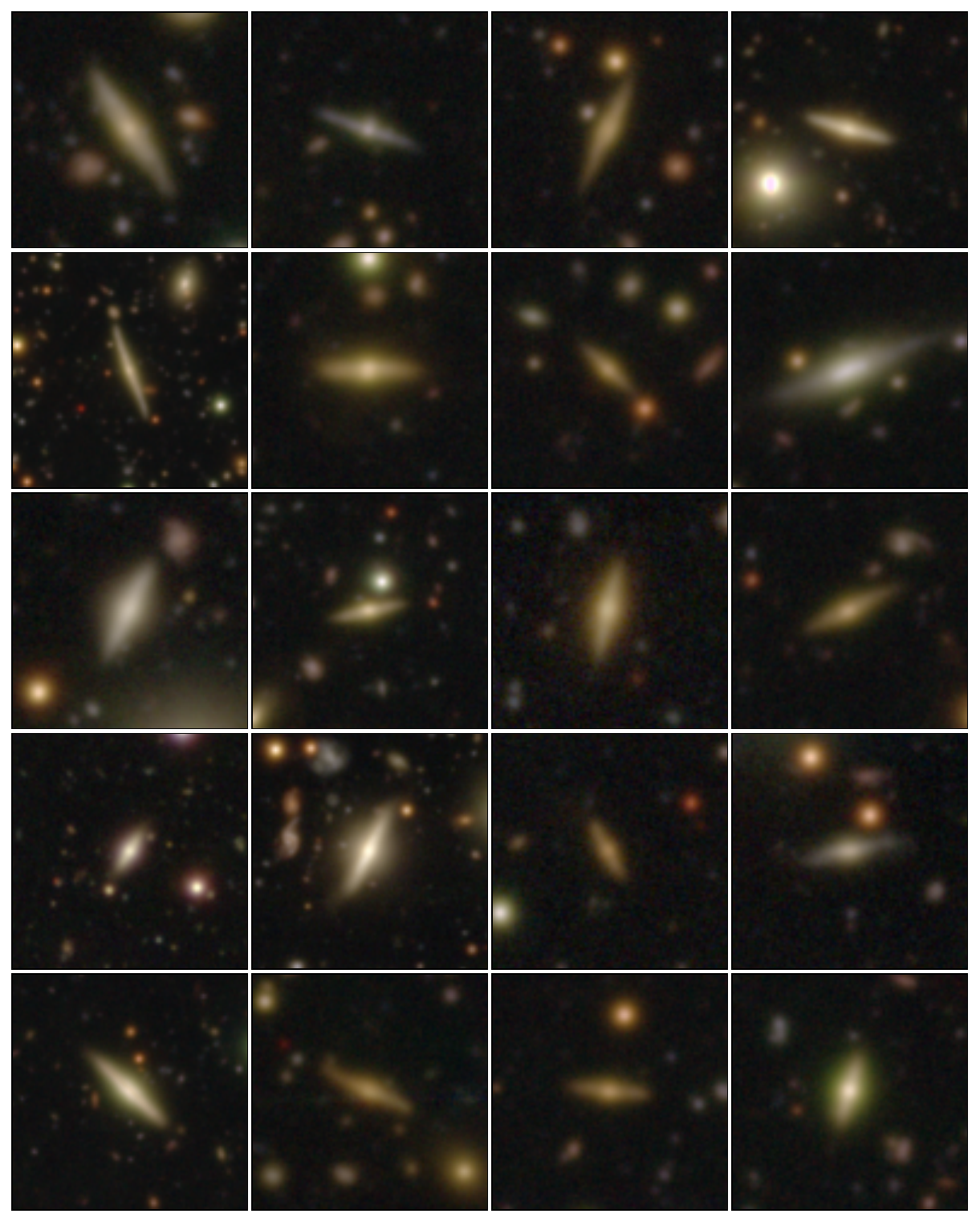}
    \caption{Example \ac{GZCD} subjects voted as being an `Edge-On Disk', having the highest volunteer vote fractions for that response following a selection cut of $f_{\rm featured-or-disk}>0.7$ for the first question. Subjects are limited to those with at least ten volunteer votes.}
    \label{fig:top20-edge-on}
\end{figure}

\begin{figure}
	\centering
	\includegraphics[width=0.9\columnwidth]{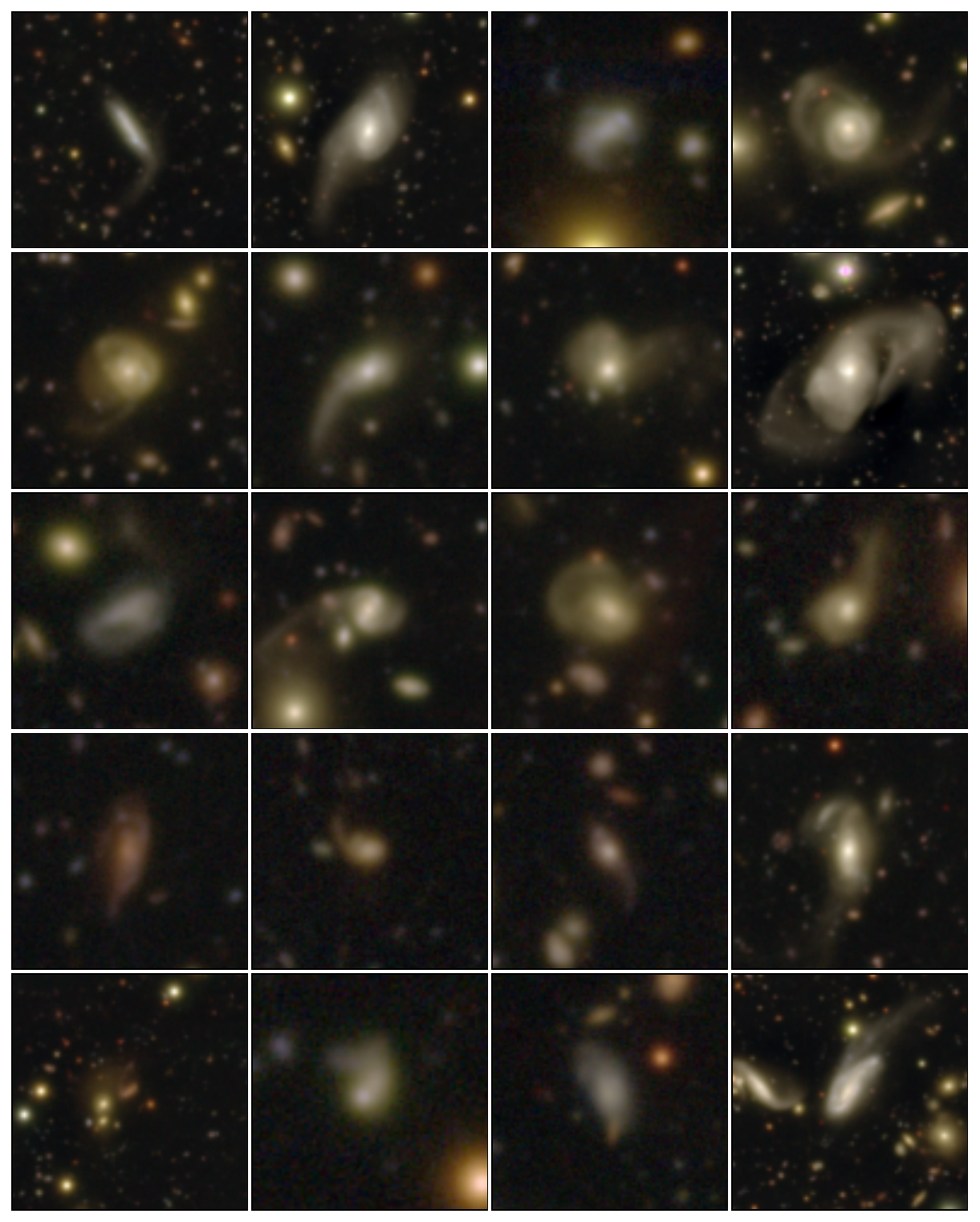}
    \caption{Example \ac{GZCD} subjects voted as being a galaxy undergoing `Major Disturbance' in response to the `Merging' question, having the highest volunteer vote fractions for that response following a selection cut of $f_{\rm problem}<0.3$ for the first question. Subjects are limited to those with at least ten volunteer votes.}
    \label{fig:top20-major-disturb}
\end{figure}

\begin{figure}
    \centering
	\includegraphics[width=0.9\columnwidth]{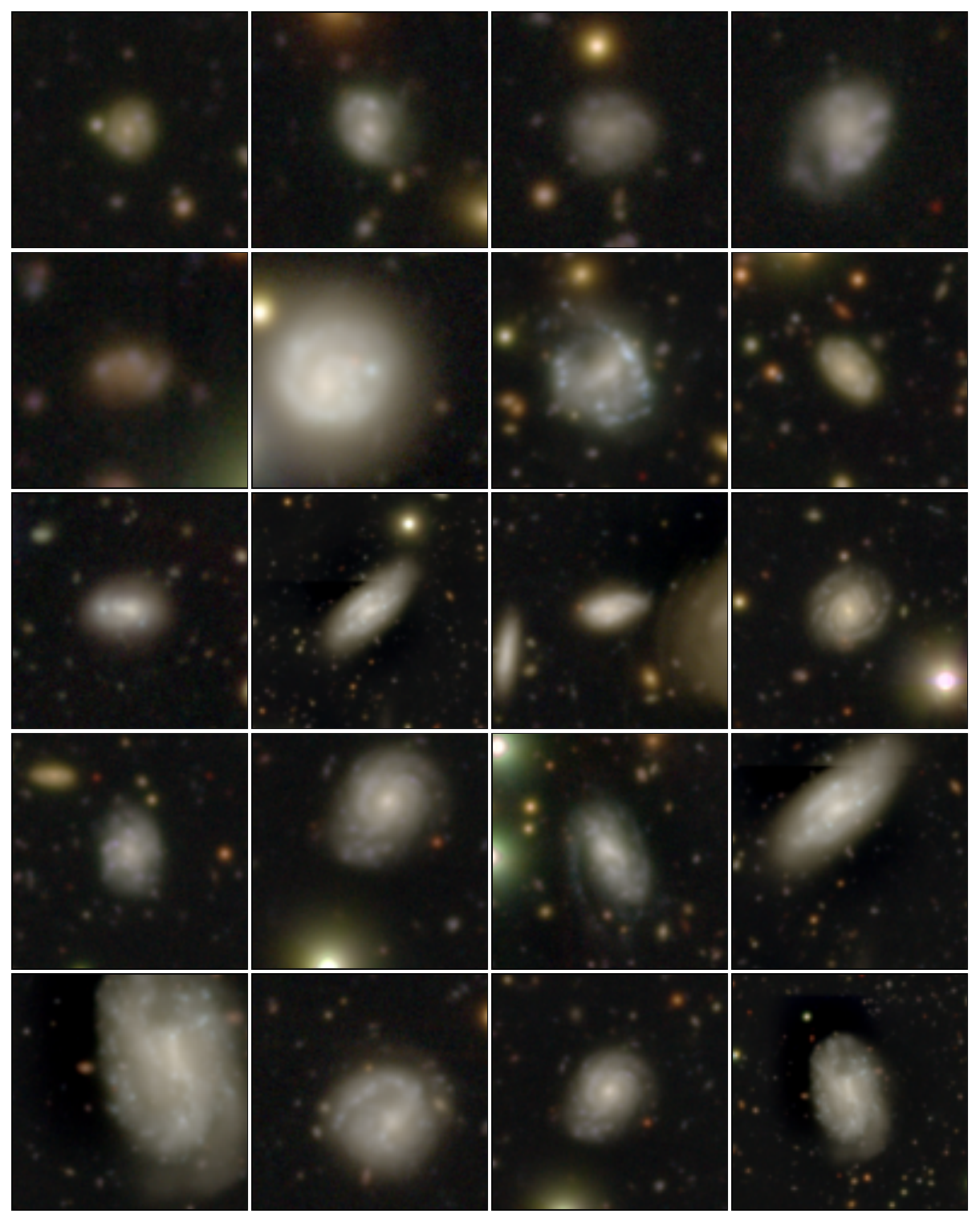}
    \caption{Example \ac{GZCD} subjects voted as being a galaxy containing `Clumps', having the highest volunteer vote fractions for that response following a selection cut of $f_{\rm featured-or-disk}>0.7$ for the first question. Subjects are limited to those with at least ten volunteer votes.}
    \label{fig:top20-clumps}
\end{figure}

\begin{figure}
	\centering
	\includegraphics[width=0.9\columnwidth]{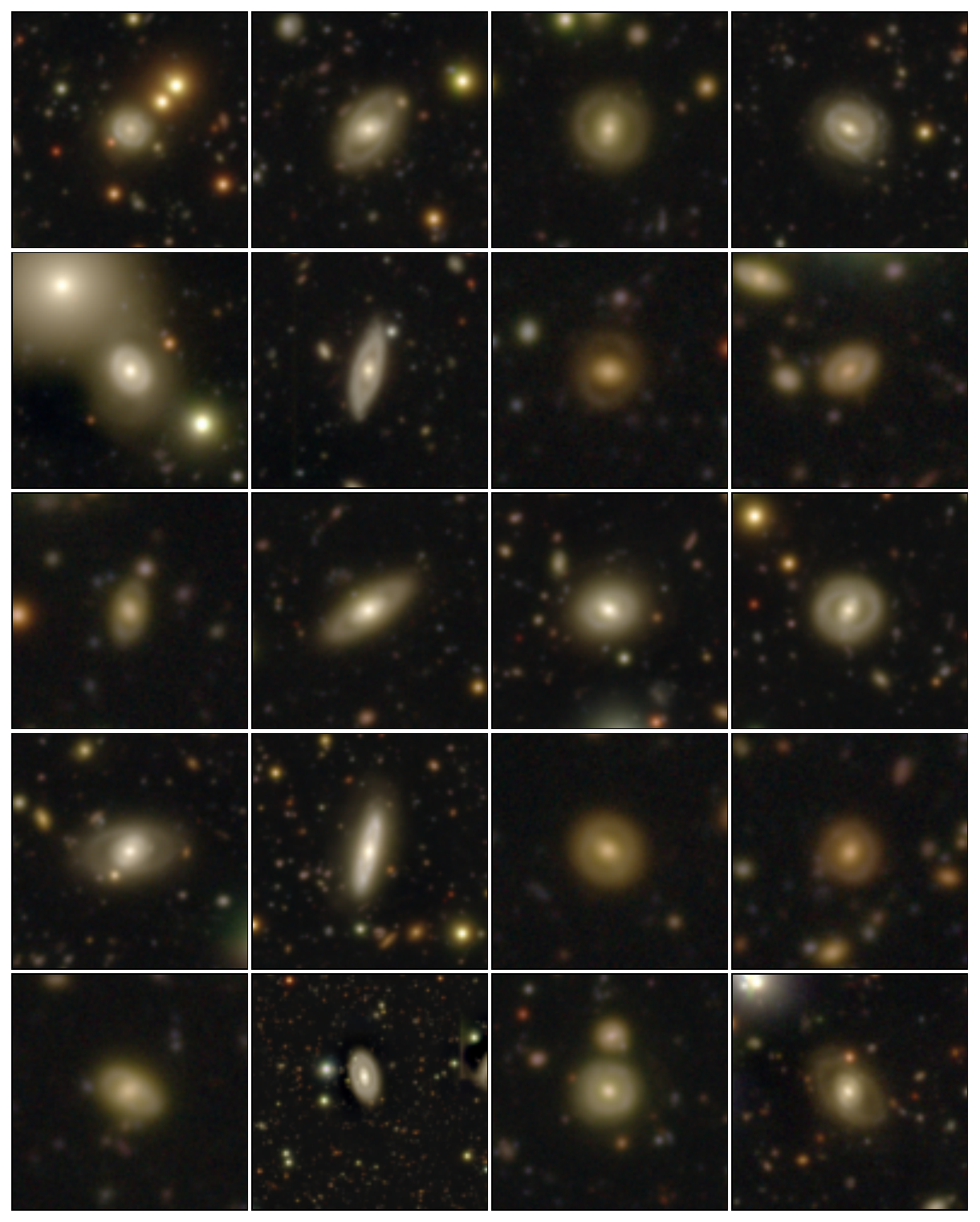}
    \caption{Example \ac{GZCD} subjects voted as being `Ring' galaxies in response to the `Rare Features' question, having the highest volunteer vote fractions for that response following a selection cut of $f_{\rm problem}<0.3$ for the first question. Subjects are limited to those with at least ten volunteer votes.}
    \label{fig:top20-ring}
\end{figure}

\begin{figure*}
	\centering
	\includegraphics[width=0.9\textwidth]{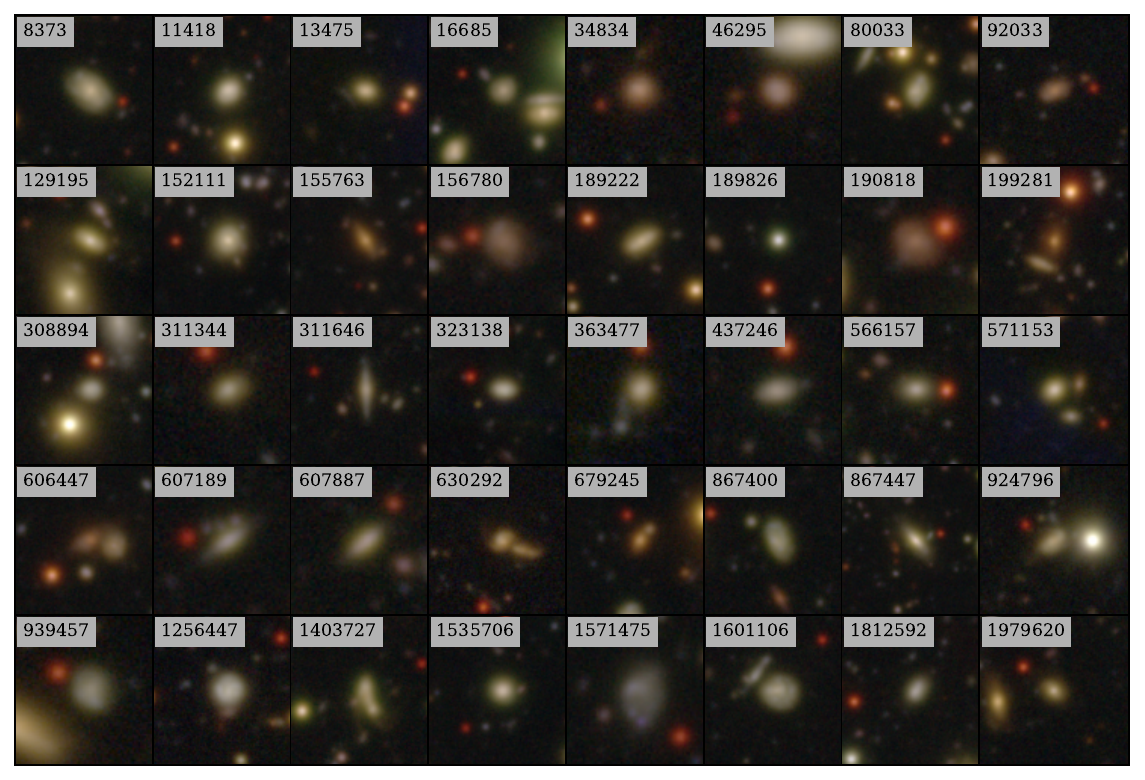}
    \caption{Example \ac{GZCD} subjects tagged by volunteers as having a ``bright-red-companion'', which appear to contain extremely red point-source objects away from the cutout centre. Subjects are listed in order of their H20 ID, which is provided next to each image.}
    \label{fig:tagged-bright-red-companion}
\end{figure*}

\begin{figure*}
	\centering
	\includegraphics[width=\textwidth]{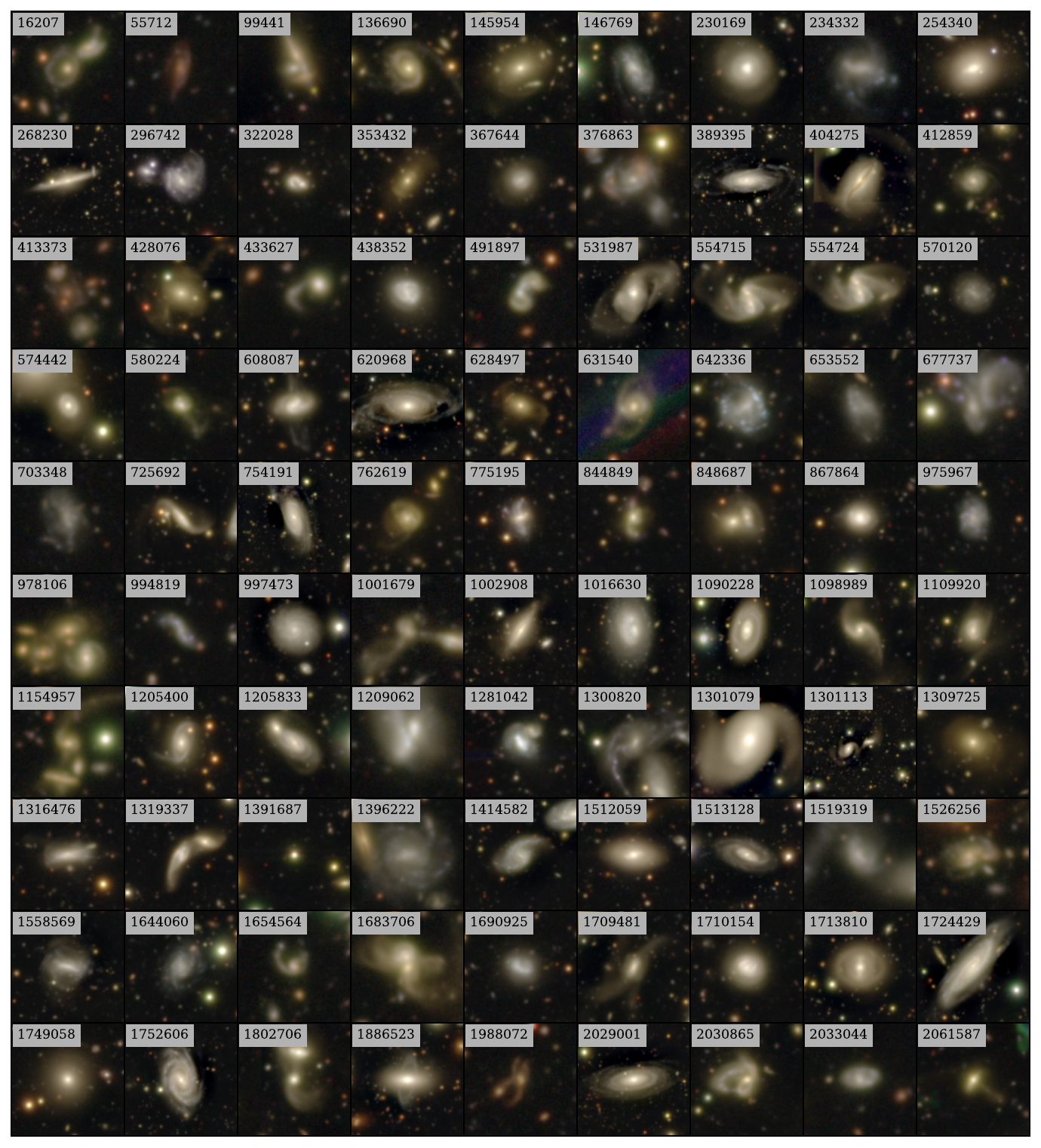}
    \caption{Example \ac{GZCD} subjects tagged by volunteers as ``beautiful'', ``interesting'', ``needs-more-research'', or similar tags. The majority correspond to spiral, irregular or disturbed galaxies with high angular resolution, as well as mergers, ring galaxies and clumpy galaxies. Subjects are listed in order of their H20 ID, which is provided next to each image.}
    \label{fig:tagged-interesting-beautiful}
\end{figure*}


\bsp	
\label{lastpage}
\end{document}